\documentclass[journal,12pt,onecolumn,draftclsnofoot,a4paper,]{IEEEtran}

\usepackage{cite}
\usepackage[T1]{fontenc}
\usepackage[latin1]{inputenc}
\usepackage{graphicx}
\usepackage{theorem}
\usepackage{amsmath}
\usepackage{amsfonts}
\usepackage{epstopdf}
\usepackage{authblk}
\usepackage[caption=false,font=footnotesize]{subfig}
\usepackage[acronym,shortcuts]{glossaries}
\usepackage{color}
\usepackage{comment}
\usepackage{cleveref}
\usepackage{xstring}
\usepackage{multirow}
\usepackage[flushleft]{threeparttable}

\newacronym{QC}{QC}{quasi-cyclic}
\newacronym{QC-LDPC}{QC-LDPC}{quasi-cyclic low-density parity-check}
\newacronym{LDPC}{LDPC}{low-density parity-check}
\newacronym{LDPCC}{LDPCC}{low-density parity-check convolutional}
\newacronym{AC-LDPC}{AC-LDPC}{array convolutional low-density parity-check}
\newacronym{PDC-LDPC}{PDC-LDPC}{progressive differences convolutional low-density parity-check}
\newacronym{SC-LDPC}{SC-LDPC}{spatially coupled low-density parity-check}
\newacronym{SC-LDPC-CC}{SC-LDPC-CC}{spatially coupled low-density parity-check convolutional code}
\newacronym{SC-LDPC-CCs}{SC-LDPC-CCs}{spatially coupled low-density parity-check convolutional codes}
\newacronym{AWGN}{AWGN}{additive white Gaussian noise}
\newacronym{BER}{BER}{bit error rate}
\newacronym{FER}{FER}{frame error rate}
\newacronym{TUB}{TUB}{truncated union bound}
\newacronym{BPSK}{BPSK}{binary phase shift keying}
\newacronym{SPA-LLR}{SPA-LLR}{sum-product algorithm with log-likelihood ratios}
\newacronym{RTI}{RTI}{regular time-invariant}
\newacronym{RTI-LDPCC}{RTI-LDPCC}{regular time-invariant low-density parity-check convolutional}
\newacronym{r.c.c.}{r.c.c.}{row-column constraint}
\newacronym{BP}{BP}{belief propagation}
\newacronym{SW}{SW}{sliding window}
\newacronym{IP}{IP}{Integer Programming}

{\theorembodyfont{\rmfamily}
   }
{\theorembodyfont{\rmfamily}
   \newtheorem{Pro}{{\textbf Proposition}}[section]}
{\theorembodyfont{\rmfamily}
   \newtheorem{Lem}{{\textbf Lemma}}[section]}
{\theorembodyfont{\rmfamily}
   \newtheorem{Cor}{{\textbf Corollary}}[section]}
{\theorembodyfont{\rmfamily}
   \newtheorem{Def}{{\textbf Definition}}[section]}
{\theorembodyfont{\rmfamily}
   \newtheorem{Exa}{{\textbf Example}}[section]}
{\theorembodyfont{\rmfamily}
   \newtheorem{Remark}{{\textbf Remark}}}

\def\HH{\mathbf{H}}
\def\PP{\mathbf{P}}

\def\0{\mathbf{0}}
\def\Hconv{\mathbf{H}}

\begin{document}


\title{Design and Analysis of Time-Invariant SC-LDPC Convolutional Codes with Small Constraint Length \thanks{A preliminary version of this paper has been presented at the IEEE BlackSea-Com Conf. 2016 \cite{BaldiBattaglioniChiaraluceEtAl2016}.} }





\author[1]{Massimo Battaglioni}
\author[2]{Alireza Tasdighi}
\author[1]{Giovanni Cancellieri}
\author[1]{Franco Chiaraluce}
\author[1]{Marco Baldi}
\affil[1]{Dipartimento di Ingegneria dell'Informazione\\
Universit\`a Politecnica delle Marche\\
Ancona, Italy\\
Email: m.battaglioni@pm.univpm.it, \{g.cancellieri, f.chiaraluce, m.baldi\}@univpm.it}
\affil[2]{Department of Mathematics and Computer Science, Amirkabir University of Technology, Tehran, Iran\\
Email: a.tasdighi@aut.ac.ir}

\renewcommand\Authands{ and }
\maketitle
\begin{abstract}
In this paper, we deal with time-invariant \ac{SC-LDPC-CCs}.
Classic design approaches usually start from \ac{QC-LDPC} block codes and exploit suitable unwrapping procedures to obtain \ac{SC-LDPC-CCs}. 
We show that the direct design of the \ac{SC-LDPC-CCs} syndrome former matrix or, equivalently, the symbolic parity-check matrix, leads to codes with smaller syndrome former constraint lengths with respect to the best solutions available in the literature.
We provide theoretical lower bounds on the syndrome former constraint length for the most relevant families of \ac{SC-LDPC-CCs}, under constraints on the minimum length of cycles in their Tanner graphs.
We also propose new code design techniques that approach or achieve such theoretical limits. 
\end{abstract}


\section{Introduction \label{sec:intro}}

\glsresetall

\Ac{SC-LDPC-CCs}\footnote{Since there is no uniformity in the literature regarding the name of the codes we consider, we follow\cite{Mitchell2015} and use the acronym \ac{SC-LDPC-CCs}.}   were first introduced
in \cite{Felstrom1999}.
Starting from \ac{QC-LDPC} block codes \cite{Lin2004Book} and applying suitable unwrapping techniques \cite{Felstrom1999, TannerBook1987,  Pusane2011}, \ac{SC-LDPC-CCs} with performance very close to the density evolution bound have been obtained. Furthermore, it has been shown that terminated \ac{SC-LDPC-CCs} asymptotically achieve Shannon's capacity \cite{Kudekar2013} under belief propagation decoding,
thanks to the threshold saturation phenomenon. 

As for other codes, an important concern of \ac{SC-LDPC-CCs} regards complexity. Implementation of \ac{SC-LDPC-CCs} can exploit shift register-based circuits for encoding \cite{Felstrom1999}, whereas \ac{SW} iterative algorithms based on belief propagation can be used for decoding \cite{Felstrom1999, Lentmaier2005, Iyengar2012}.
Through these solutions, the encoding and decoding complexity increases linearly with the product of two parameters of the code: the syndrome former constraint length and the weight of the columns of the parity-check matrix. So, minimization of the syndrome former constraint length is a relevant goal from the complexity standpoint.

The performance and complexity of these codes also depend on the approach followed for their design. As mentioned above, \ac{SC-LDPC-CCs} are often obtained starting from \ac{QC-LDPC} block codes and then applying unwrapping techniques. Freedom in the code design can be increased by avoiding this intermediate step and designing \ac{SC-LDPC-CCs} directly. Indeed, such an approach has already been followed in \cite{Baldi2012b,Cho2015} for codes having rates $R=\frac{a-1}{a}$,
and extended in \cite{BaldiBattaglioniChiaraluceEtAl2016} to codes with rate $R=\frac{a-c}{a}$, where $a$ and $c<a$ are integers.
However, to the best of our knowledge, a general and exhaustive study is still missing.

On the other hand, \ac{SC-LDPC-CCs} can be also designed by means of random, algebraic or combinatorial design methods \cite{Chandrasetty2014, Liu2016, Zhang2016}; codes designed this way can exhibit excellent performance in the error floor region, i.e., the region corresponding to high values of the signal-to-noise ratio (SNR). An important role, in this region, is played by the code minimum distance.
Another feature that significantly affects performance of these codes under iterative decoding is the length of cycles in their associated Tanner graphs \cite{Gallager,Tanner1981}. The performance of belief propagation decoders is adversely affected by the presence of cycles with short length and, in particular, with length 4 in the code Tanner graph.
Therefore, code design approaches often aim at increasing the minimum length of cycles, also known as girth of the graph. The relationship between girth and minimum distance has been deeply investigated in \cite{Fossorier2004, Smarandache2012}.

Unfortunately, good girth profiles may not coincide with optimal minimum distance properties. This is due to the impact of absorbing sets, a particular type of trapping sets \cite{Vasic2011} that originate from clusters of short cycles in the Tanner graph. Design approaches based on the exact enumeration of absorbing sets have been proposed  \cite{Amiri2015}. Nevertheless, it has been demonstrated that a punctual removal of some structures of cycles is usually a sufficient condition for the removal of the most harmful absorbing sets \cite{Tao2017}. So, in order to have good performance in the high SNR region, a possible approach consists in removing all the cycles with a given short length \cite{Wang2008}. This approach has been shown to achieve good results \cite{Wang2013}. Motivated by the above considerations, we focus our optimization on the girth properties, which require a simpler analytical investigation and, eventually, lead to a good performance in the error-floor region.

The main goal of this paper is to show that \ac{SC-LDPC-CCs} with small syndrome former constraint length can be effectively designed, which are characterized by good error rate performance and, most of all, reduced complexity with respect to previous solutions. For such a purpose, we introduce a new direct design approach that does not need unwrapping of a block code. The proposed method uses the symbolic parity-check matrix that, in the form commonly considered in the literature \cite{Tasdighi2016, Fossorier2004, BocharovaHugJohannessonEtAl2012}, has some fixed entries. We show that the removal of this constraint makes investigations easier and yields significant reductions in terms of constraint length.

The rationale of our approach is a careful analysis of the trade-off between good girth properties and small constraint length. More precisely, we derive theoretical lower bounds on the syndrome former constraint length for many families of \ac{SC-LDPC-CCs}, under constraints on the girth, and we provide several explicit examples of codes that are able to approach or even to reach these bounds.

The remainder of the paper is organized as follows. 
In Section \ref{sec:TICodes} we recall the definitions of time-invariant \ac{SC-LDPC-CCs} and introduce the basic notation used to represent these codes and their relevant features.
In Section \ref{sec:MinConstLength} some bounds on the syndrome former constraint length of a wide range of code families and several girth lengths are provided. In Section \ref{sec:Methods} we introduce three new design methods for as many families of time-invariant \ac{SC-LDPC-CCs} with short syndrome former constraint length. In Section \ref{sec:Examples} we perform heuristic code searches and compare their results with the theoretical bounds. We also assess and compare the error rate performance of the designed codes. Finally, in Section \ref{sec:Conclusion} we draw some conclusions.

\section{Notation and definitions \label{sec:TICodes}}

Let us introduce the notation and definitions that will be used throughout the paper.

\subsection{
Time-invariant \ac{SC-LDPC-CCs} \label{sec:TILDPCCC}}

Time-invariant \ac{SC-LDPC-CCs} are characterized by semi-infinite parity-check matrices in the form
\begin{equation}
\HH = \left[\begin{array}{cccccc}
\arraycolsep=1.4pt\def\arraystretch{5pt}
\HH_0 				& \0 							& \0 							& \ddots \\
\HH_1				& \HH_0					& \0 							& \ddots \\
\HH_2				& \HH_1					& \HH_0					& \ddots \\
\vdots 			& \HH_2 				& \HH_1 				& \ddots \\
\HH_{m_h}	& \vdots 				& \HH_2					& \ddots \\
\0 						& \HH_{m_h}	& \vdots					& \ddots \\
\0 						& \0 							& \HH_{m_h}	& \ddots \\
\0 						& \0 							& \0							& \ddots \\
\vdots				& \vdots					& \vdots					& \ddots \\
\end{array}\right],
\label{eq:Hconv}
\end{equation}
where each block $\HH_i$, $i = 0, 1, 2, \ldots, m_h$, is a binary matrix with size $c \times a$, that is,
\begin{equation}
\HH_i=\left[\begin{array}{llll}
h_{i}^{(0,0)} & h_{i}^{(0,1)} & \ldots & h_{i}^{(0,a-1)}\\
h^{(1,0)}_{i} & h_{i}^{(1,1)} & \ldots &h_{i}^{(1,a-1)}\\
\vdots & \vdots & \ddots & \vdots\\
h^{(c-1,0)}_{i} &h_{i}^{(c-1,1)} & \ldots & h^{(c-1,a-1)}_{i}\end{array}\right],
\label{eq:Hi}
\end{equation}
where each $h_{i}^{(j,k)}$, $j = 0, 1, 2, \ldots, c-1$, $k = 0, 1, 2, \ldots, a-1$, is a binary entry.
The syndrome former matrix $\HH_s = \left[ \HH_0^T | \HH_1^T | \HH_2^T | \ldots | \HH_{m_h}^T \right]$, where the superscript $T$ denotes transposition, has $a$ rows and $(m_h+1)c$ columns. Let us introduce the variable $L_h$, which is defined as the maximum spacing between any two ones appearing in any column of $\HH_s^T$.
The code is column-regular if all the columns of $\HH$ have the same Hamming weight $w_c$. Similar arguments hold for row-regular codes \footnote{Even if the code is row-regular, the first and the last $m_hc$ rows have a lower weight with respect to the central ones. This is due to the initial and final terminations of the code.}. If the code is regular in both its columns and rows, it is said to be regular. In this paper row regularity is not a requirement; so, with a small abuse of notation, we will use interchangeably the terms column-regular and regular and, similarly, the terms column-irregular and irregular.

As evident from (\ref{eq:Hconv}), $\HH$ is obtained by $\HH_s^T$ and its replicas, shifted vertically by $c$ positions each. The code defined by \eqref{eq:Hconv} has asymptotic code rate $R = \frac{a-c}{a}$, syndrome former memory order $m_h = \left\lceil \frac{L_h}{c} \right\rceil - 1$, where $\left\lceil x \right\rceil$ is the smallest integer greater than or equal to $x$, and syndrome former constraint length $v_s = (m_h + 1) a = \left\lceil \frac{L_h}{c} \right\rceil a$.

The symbolic representation of $\HH_s$
exploits polynomials $\in F_2[x]$, where $F_2[x]$ is the ring of polynomials with coefficients in the binary Galois field $\mathrm{GF}[2]$. In this case, the code is described by a $c \times a$ \textit{symbolic matrix} having polynomial entries, 
that is
\begin{equation}
H(x)=\left[\begin{array}{llll}
h_{0,0}(x) & h_{0,1}(x) & \ldots & h_{0,a-1}(x)\\
h_{1,0}(x) & h_{1,1}(x) & \ldots & h_{1,a-1}(x)\\
\vdots & \vdots & \ddots & \vdots\\
h_{c-1,0}(x) & h_{c-1,1}(x) & \ldots & h_{c-1,a-1}(x)\end{array}\right],
\label{eq:Hx}
\end{equation}
where each $h_{i,j}(x)$, $i = 0, 1, 2, \ldots, c-1$, $j = 0, 1, 2, \ldots, a-1$, is a polynomial $\in F_2[x]$.
The code representation based on $\HH_s$ can be converted into that based on $H(x)$ as follows
\begin{equation}
h_{i,j}(x)=\sum_{m=0}^{m_h} h_{m}^{(i,j)}\cdot x^{m}.
\label{eq:bintopol}
\end{equation}

The syndrome former memory order $m_h$ coincides with the largest difference, in absolute value, between any two exponents of the variable $x$ appearing in the polynomial entries of $H(x)$.
The highest weight of any polynomial entry of $H(x)$ defines the type of the code. 
So, a code containing only monomial or null entries is called here monomial or Type-$1$ code, a code having also binomial entries is called binomial or Type-$2$ code and so on. In this paper, up to Type-$3$ codes are considered in the design of the symbolic parity-check matrix. The matrix $H(x)$ can be described by the exponents of $x$ without loss of information. This permits us to define the exponent matrix $\PP$ of Type-$1$ codes such that $p_{i,j}=\log_x(h_{i,j}(x))$; conventionally, $\log_x(0)$ is denoted as $-1$.

Most previous works are devoted to the design of $H(x)$, from which $\HH_s$ is obtained by unwrapping \cite{Felstrom1999, TannerBook1987,  Pusane2011}.
However, designing $H(x)$ requires first to choose the form of the polynomials $h_{i,j}(x)$ (null, monomials, binomials, etc.) and then optimize their exponents.
The matrix $H(x)$ is also used in \cite{Zhou2012} to find unavoidable cycles and design monomial \ac{SC-LDPC-CCs} free of short cycles.
On the other hand, working with $\HH_s$ is advantageous as it allows to perform a single step optimization over all possible choices.
As we will see in Sections \ref{sec:Methods} and \ref{sec:Examples}, such an approach permits us to find codes with shorter constraint length with respect to previous approaches.

\subsection{
Cycles Characterization \label{sec:LocalCycles}}

Cycles are closed loops starting from a node of the Tanner graph associated to a code and returning to the same node without passing more than once through any edge.
Since the parity-check matrix is the bi-adjacency matrix of the Tanner graph of a code, cycles can be defined over such a matrix as well. 
This way, we are able to directly relate the syndrome former constraint length of a \ac{SC-LDPC-CC} to its cycles length.

A common constraint in Tanner graphs of \ac{LDPC} codes is that their girth must be  greater than $4$, that is, cycles with length $4$ must not exist in the code parity-check matrix; if this condition is true, the so-called \ac{r.c.c.} is satisfied \cite{RyanBook}.

Following an approach similar to that introduced in \cite{Baldi2012b}, we describe the matrix $\HH_s^T$ through a set of integer values representing the position differences between each pair of ones in its columns.
A similar approach based on differences has also been adopted in \cite{Fossorier2004} for \ac{QC-LDPC} block codes and their symbolic matrices.

\begin{Def}
We denote the position difference between a pair of ones in a column of $\HH_s^T$ as $\delta_{i,j}$, where $i$ is the column index of $\HH_s^T$ $(i=0,1,2,\ldots,a-1)$ 
and $j$ is the row index of $\HH_s^T$ corresponding to the first of the two symbols $1$ involving the difference $(j\in [0,1,2,\ldots, L_h-2])$.
The index of the second symbol $1$ involved in the difference is simply found as $j + \delta_{i,j}$.
For each difference, we also compute the values of two \textit{levels} which are relative to the parameter $c$.
The \textit{starting level} is defined as $s_l = j \mod c$, while the \textit{ending level} is defined as $e_l = (j + \delta_{i,j}) \mod c$.
Both levels obviously take values in $\left\{0,1,2 \ldots, c-1\right\}$.
\label{def:differences}
\end{Def}

\begin{figure}[t]
\begin{centering}
\includegraphics[width=80mm,keepaspectratio]{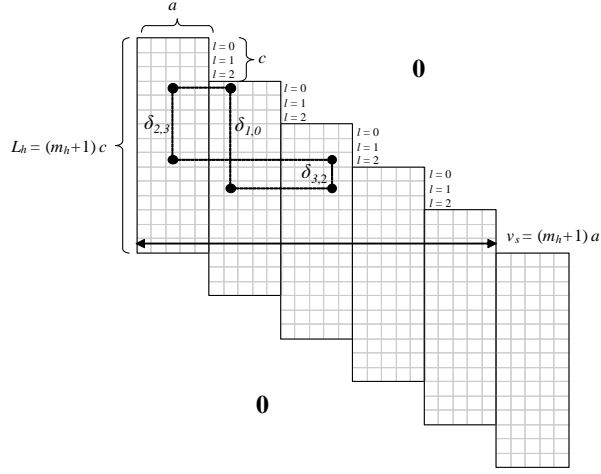}
\caption{Example of (a section of) $\Hconv$ with a cycle of length $6$.}
\label{fig:CycleExample}
\par\end{centering}
\end{figure}

Based on this representation of $\HH_s^T$, it is easy to identify closed loops in the Tanner graph associated to $\Hconv$.
In fact, a cycle is associated to a sum of the type $\delta_{i_1,j_1} \pm \delta_{i_2,j_2} \pm \ldots \pm \delta_{i_L,j_L}=0$,
and the length of the cycle is $2L$, with $L$ being an integer $> 1$. 
An example is reported in Fig. \ref{fig:CycleExample}, where a cycle with length $6$ corresponds to the relation $\delta_{2,3} + \delta_{3,2} - \delta_{1,0} = 0$, in a code with $a=5$, $c=3$ and $m_h=4$. For the sake of simplicity, the figure assumes that $L_h$ is a multiple of $c$.

Not all the possible sums or differences of $\delta_{i,j}$ generate cycles.
In fact, $\delta_{i_1,j_1}$ can be added to $\delta_{i_2,j_2}$ in order to form a cycle if and only if the starting level of the latter coincides with the ending level of the former.
Instead, $\delta_{i_1,j_1}$ can be subtracted to $\delta_{i_2,j_2}$ in order to form a cycle if and only if their ending levels coincide.
In addition, the starting level of the first difference and the starting level of the last difference in 	$\delta_{i_1,j_1} \pm \delta_{i_2,j_2} \pm \ldots - \delta_{i_L,j_L}$ must coincide. Similarly, the starting level of the first difference and the ending level of the last difference in 	$\delta_{i_1,j_1} \pm \delta_{i_2,j_2} \pm \ldots + \delta_{i_L,j_L}$ must coincide.
Let us denote as $\left. \delta_{i,j}\right.^{(s_l\rightarrow e_l)}$ the difference $\delta_{i,j}$ with its starting and ending levels $s_l$ and $e_l$.
In the example reported in Fig. \ref{fig:CycleExample}, we have $\left. \delta_{2,3} \right.^{(0\rightarrow 2)} + \left. \delta_{3,2} \right.^{(2\rightarrow 1)} - \left. \delta_{1,0} \right.^{(0\rightarrow 1)} = 0$,
which therefore complies with the above rules.
Based on these considerations, for a given $\HH_s^T$ an efficient numerical procedure can be exploited to find all the
cycles within a given maximum length.
Moreover, by studying the cases in which differences can or cannot be summed or subtracted, it is possible to obtain lower bounds on the constraint length which is needed to avoid cycles up to a given length, as described in the next
sections.

\subsection{Symbolic parity-check matrix representation \label{sec:Symbolic}}

Bipartite graphs $\tilde{G}_1\left(\tilde{U}_1 \cup \tilde{W}_1, \tilde{E}_1\right)$ and $\tilde{G}_2\left(\tilde{U}_2 \cup \tilde{W}_2, \tilde{E}_2\right)$ are isomorphic if there is a bijection $f: \tilde{U}_1 \cup \tilde{W}_1 \rightarrow \tilde{U}_2 \cup \tilde{W}_2$ such that $e_1 = \left\{u_1, w_1 \right\}$ is an element of $\tilde{E}_1$ if and only if $e_2 = \left\{f\left(u_1\right), f\left(w_1\right) \right\}$ is an element of $\tilde{E}_2$. Equivalently, if $\tilde{\HH}_1$ and $\tilde{\HH}_2$ are respectively bi-adjacency matrices of isomorphic bipartite graphs $\tilde{G}_1$ and $\tilde{G}_2$ we say that $\tilde{\HH}_1$ and $\tilde{\HH}_2$ are graph-isomorphic if there is a bijection $f$ so that entry $\left(\tilde{\HH}_1\right)_{ij} = 1$ if and only if $\left(\tilde{\HH}_2\right)_{f(i)f(j)} = 1$; the domain of the bijection is the union of the row labels and the column labels of the parity-check matrix $\HH_1$ \cite{Tasdighi2016}. If two parity-check matrices are graph-isomorphic, then also their corresponding exponent matrices are graph-isomorphic. Graph-isomorphic parity-check matrices define equivalent codes ($C_1\cong C_2$). Equivalent codes have the same minimum distance, girth and performance.

In the following we provide three lemmas defining useful properties of exponent matrices of Type-$1$ \ac{QC-LDPC} codes. Their proofs can be found in \cite{Tasdighi2016}. These lemmas also hold for \ac{SC-LDPC-CCs}.

\begin{Lem}
Let $\PP_1$ and $\PP_2$ be the exponent matrices of the codes $C_1$ and $C_2$. If $\PP_1$ can be obtained by permuting the rows or the columns of $\PP_2$, then $C_1\cong C_2$.
\label{lem:permequi}
\end{Lem}

\begin{Lem}
Let $\PP_1$ and $\PP_2$ be the exponent matrices of the codes $C_1$ and $C_2$. If $\PP_1$ can be obtained by adding or subtracting the same constant to all the elements of a row or a column of $\PP_2$, then $C_1\cong C_2$.
\label{lem:sumequi}
\end{Lem}

\begin{Lem}
Let $\PP_1$ and $\PP_2$ be the exponent matrices of the codes $C_1$ and $C_2$. Also, let $p$ be the size of the circulant permutation matrices in $\HH_i\;(i=1,2)$. Assume that $d\in \lbrace1,2,...,p-1\rbrace$ and $p$ are co-prime. If $(\PP_2)_{ij}=(d (\PP_1)_{ij})\mod p$ for $0\leq i\leq c-1$ and $0\leq j\leq a-1$, then $C_1\cong C_2$.
\label{lem:proequi}
\end{Lem}

\begin{Remark}
The parity-check matrix of a tail-biting \ac{SC-LDPC-CC} \cite{Tavares2007} with blocklength $a p$ and rate $\frac{a-c}{a}$ can be obtained by properly reordering the rows and columns of the parity-check matrix of an equivalent \ac{QC-LDPC} block code, which is a $c \times a$ array of $p \times p$ circulant permutation matrices. When $p \rightarrow \infty$, the effect of the tail-biting termination becomes negligible, and an \ac{SC-LDPC-CC} is obtained.
\end{Remark}

The above lemmas permit us to design w.l.o.g. the exponent matrices of \ac{SC-LDPC-CCs} with at least a null entry in any row and any column. This is because if there is a row (or a column) such that all the exponents are greater than $0$, it is always possible (by using Lemma \ref{lem:sumequi}) to subtract the minimum among the exponents to all the elements of the row (or column), thus obtaining at least a null entry.

\subsubsection*{IP model \label{subsec:IP}}

We also propose to use an optimization \ac{IP} model \cite{Wolsey1998} in order to find the minimum possible $m_h$ for each exponent matrix of monomial codes with $w=c$. We call this method Min-Max; its description is given in Appendix A. This model, instead of performing an exhaustive search, takes benefit of a heuristic optimization approach. This process significantly decreases the search time. The input exponent matrices of our Min-Max model were selected from \cite{Tasdighi2016,BocharovaHugJohannessonEtAl2012,Tasdighi2016a}.

\section{Lower bounds on the constraint length \label{sec:MinConstLength} }

We consider some specific families of \ac{SC-LDPC-CCs} and aim at estimating the minimum syndrome former
constraint length which is needed to achieve a certain girth $g$.

For the sake of convenience, according to the definition given in Section \ref{sec:TICodes}, we distinguish between the case of Type-$1$ codes and Type-$z$ codes, with $z >1$. For the latter case, though most of the results obtained are valid for any $z$, in the following we will focus on the cases $z = 2$ and $z = 3$, which represent a good trade-off between performance and complexity.

The bounds are described in terms of $m_h$ when the analysis is performed on $H(x)$ and in terms of $L_h$ when $\HH_s^T$ is studied.
We remind that these parameters are directly related to the syndrome former constraint length $v_s$.
So, the bounds on $v_s$ can be easily computed as well.

\renewcommand*{\thefootnote}{\fnsymbol{footnote}}

The bounds are presented and proved in the following subsections. For the sake of clarity, the main results are summarized in Table \ref{Table:lowboundstab}.

\begin{table}[ht]
\renewcommand{\arraystretch}{1}
\small
\centering
\begin{threeparttable}
\caption{Lower bounds on the constraint length.}
\label{table:GT}
\begin{tabular}{|c|c|c|c|c|}
\hline
Lower bound & Family & $g$ & $w$ & $c$\\
\hline
\hline
$m_h\geq \Big\lceil \frac{a-1}{2} \Big\rceil$ & Type-$1$ & $\geq 6$ & any  & $w$ \\ [0.5ex] 
\hline
$m_h\geq \Big\lceil \frac{a-1}{2} \Big\rceil$ & Type-$1$ & $\geq 8$ & 2  & $2$ \\
\hline
$m_h\geq \Big\lceil \frac{a(a-1)}{8} \Big\rceil$ & Type-$1$ & $\geq 8$ & 3  & 3 \\
\hline
$ m_h \ge \frac{3}{2} \binom{a}{2}$ & Type-$1$ & $\geq10$ & 3 & 3\\
\hline
$ m_h \ge \frac{1}{2} \binom{a}{2}$ & Type-$1$ & $12$ & 2 & 2\\
\hline
$ m_h \ge a-1$ & Type-1c \tnote{*} & $\geq 6$ & 3 & 3\\
\hline
$ m_h \ge \frac{a(c-1)}{2}$ & Type-1c \tnote{*} & $\geq 8$ & 3 & 3\\
\hline
$L_h \ge  \Big\lceil \frac{ \sum_{i=0}^{a-1}\binom{w_i}{2}+\binom{c+1}{2}}{c} \Big\rceil$& Type-$z$ & $\geq 6$ & any & any\\
\hline
$L_h \ge  \Big\lceil \frac{2 \sum_{i=0}^{a-1}\binom{w_i}{2}}{c} \Big\rceil $& Type-$z$ & $\geq 8$ & any & any\\
\hline
\end{tabular}
\label{Table:lowboundstab}
\begin{tablenotes}
      \small
      \item[*] Type-1c codes are characterized by a symbolic matrix having only ones in the first row.
    \end{tablenotes}
\end{threeparttable}
\end{table}

\subsection{Type-1 codes}

\subsubsection{Lower bounds on $m_h$}

Type-1 codes are characterized by a symbolic form of the parity-check matrix containing only monomial or null entries. If $\HH_s^T$ contains a column with weight $w>c$, the code cannot be a monomial code because the number of entries exceeds the number of rows of $H(x)$.
Therefore, in this paper we will focus on the case $w=c$.

The following lemma provides a necessary condition on $m_h$ for fulfilling the \ac{r.c.c.}. Longer girths are studied next.

\begin{Lem}
A Type-1 code fulfilling the \ac{r.c.c.} and having $g \geq 6$, $\forall w=c$, has
\begin{equation}
m_h\geq \Big\lceil \frac{a-1}{2} \Big\rceil.
\label{eq:LhBoundCycles4w3monomial}
\end{equation}
\label{lem:lemmaonmonsw3}
\end{Lem}
\begin{IEEEproof}
The difference between any two exponents in a column of $H(x)$ takes values in $\left[-m_h, m_h\right]$. Any difference must appear no more than once to meet the \ac{r.c.c.}.
Exploiting all values in $\left[-m_h, m_h\right]$ to design the matrix columns, we obtain $a=2m_h+1$, from which (\ref{eq:LhBoundCycles4w3monomial}) is derived.
\end{IEEEproof}

A very important property of monomial codes with parity-check matrix column weight $w=c=2$ is that cycles with length $2(2k+1)$, $k=1,2,\ldots,$ cannot exist because of structural characteristics.
In fact, in monomial codes with $c=w=2$, any difference starting from the first level ends in the second level, and vice versa; this means that \begin{equation}
s_{li} \neq e_{li} \quad \forall i.
\label{eq:diffsl}
\end{equation}
In order to form a cycle with length $2L$, the starting level of the first difference and the starting level of the last difference in $\delta_{i_0,j_0}^{(s_{l0}\rightarrow e_{l0})} \pm \delta_{i_1,j_1}^{(s_{l1}\rightarrow e_{l1})} \pm \ldots - \delta_{i_L,j_L}^{(s_{lL}\rightarrow e_{lL})}=0$ must coincide. Similarly, the starting level of the first difference and the ending level of the last difference in $\delta_{i_0,j_0}^{(s_{l0}\rightarrow e_{l0})} \pm \delta_{i_1,j_1}^{(s_{l1}\rightarrow e_{l1})} \pm \ldots + \delta_{i_L,j_L}^{(s_{lL}\rightarrow e_{lL})}=0$ have to be the same. If $L=2k+1$, (\ref{eq:diffsl}) forces $s_{l0} \neq s_{l(2k+1)}$ in the first case, and $s_{l0} \neq e_{l(2k+1)}$ in the second case, thus preventing the occurrence of the cycle.

Therefore, cycles of length $6$ cannot exist for $w = c= 2$.

In order to find a lower bound on $m_h$ for $w = c = 3$, we can proceed as follows.
Let us consider $H(x)$ and its corresponding exponent matrix $\PP$. From the latter we can compute the matrix $\Delta \PP$, where $\Delta p_{0,j}=p_{1,j}-p_{0,j}$, $\Delta p_{1,j}=p_{2,j}-p_{1,j}$, $\Delta p_{2,j}=p_{2,j}-p_{0,j}$. Obviously, $\Delta p_{0,j}+\Delta p_{1,j}=\Delta p_{2,j}$. To satisfy the \ac{r.c.c.} it must be $\Delta p_{i,j_1} \neq \Delta p_{i,j_2}$, $\forall i$, $\forall j_1 \neq j_2$. It is proved in \cite{Fossorier2004} that a cycle of length six must involve three columns and three rows of $H(x)$, this means that the equality $\Delta p_{0,j_1}+\Delta p_{1,j_2}=\Delta p_{2, j_3}$ covers all possible length-$6$ cycles.

\begin{Lem}
The elements of $\Delta \PP$ for a monomial code with $w = c = 3$ whose Tanner graph has $g\geq 8$ satisfy the following properties:
\begin{enumerate}
\item{$\Delta p_{i,j}\in [-m_h,m_h]$, $\forall i,j$}
\item{$\Delta p_{i,j_1}\neq \Delta p_{i,j_2}$, $\forall i$, $\forall j_1 \neq j_2$}
\item{$\Delta p_{0,j_1}+\Delta p_{1,j_2}\neq \Delta p_{2,j_3}$, $\forall j_1\neq j_2,j_3$, $\forall j_2\neq j_3$}
\end{enumerate}
\label{lem:mats}
\end{Lem}

\begin{IEEEproof}
\begin{enumerate}
\item Any $p_{i,j}\in [0, m_h]$, $\forall i,j$, by definition. Since any $\Delta p_{i,j}$ is a difference between two entries of $\PP$, it can take values in $[-m_h,m_h]$.
\item In order to satisfy the \ac{r.c.c.}, it must be $p_{i_1,j_1}-p_{i_2,j_1}\neq p_{i_1, j_2}-p_{i_2,j_2}$, $\forall i_1 \neq i_2$, $\forall j_1 \neq j_2$, that is, $\Delta p_{i,j_1}\neq \Delta p_{i,j_2}$, $\forall i$, $\forall j_1 \neq j_2$.
\item A cycle with length $6$ occurs if and only if an equation of the type
\[
(p_{1,j_1}-p_{0,j_1})+(p_{2,j_2}-p_{1,j_2})=(p_{2,j_3}-p_{0,j_3})
\]
is verified, for some $j_1, j_2, j_3$, with $j_1\neq j_2, j_3$ and $j_2\neq j_3$. 
Hence, for a given $j_3$, this may occur $\forall j_2\neq j_3$ and $\forall j_1\neq j_2, j_3$.
\end{enumerate}
\end{IEEEproof}

\begin{Lem}

A necessary condition to have $g \geq 8$ in monomial codes with $w=c=3$ is
\begin{equation}
m_h\geq \Bigg\lceil \frac{a(a-1)}{8} \Bigg\rceil.
\label{eq:lemg8w3mon}
\end{equation}
\end{Lem}

\begin{IEEEproof}
Let us consider a generic entry in the third row of $\Delta \PP$, namely $\Delta p_{2,j}$.
In order to satisfy condition $3)$ of Lemma \ref{lem:mats}, $\Delta p_{2,j_3}$ must be different from all the sums of an element $\Delta p_{0,j_1}$, $\forall j_1 \neq j_3$ with an element $\Delta p_{1,j_2}$, $\forall j_2 \neq j_1, j_3$. Furthermore, for condition 2) of Lemma \ref{lem:mats}, $\Delta p_{2,j_1}=\Delta p_{0,j_1} +\Delta p_{1,j_1} \neq \Delta p_{2,j_3}$, $\forall j_1 \neq j_3$. Thus, there must be $\sum_{i=0}^{a-2}{(a-i-1)}=\binom{a}{2}$ different sums which are also different from $\Delta p_{2,j}$. Since $(\Delta p_{0,j_1} +\Delta p_{1,j_2}) \in [-2m_h,2m_h]$, the different sums can assume $4m_h+1$ values (all values except $\Delta p_{2,j}$). It follows that
\begin{equation}
\binom{a}{2}\leq 4m_h,
\end{equation}
from which (\ref{eq:lemg8w3mon}) is obtained. 
\end{IEEEproof}

\begin{Def}
We define a generic difference $\Delta p_{j}^{(i_1 \leftrightarrow i_2)} = p_{i_1,j}-p_{i_2,j}$. A difference of differences is defined as 
\begin{equation}
\begin{split}
&\Delta p_{j_1}^{(i_1\leftrightarrow i_2)} - \Delta p_{j_2}^{(i_1\leftrightarrow i_2)} = p_{i_1,j_1}-p_{i_2,j_1} -p_{i_1,j_2}+p_{i_2,j_2}, \\ & \forall  j_1 \neq j_2, i_1 \neq i_2.
\end{split}
\label{eq:gendiffs}
\end{equation}
\end{Def}

\begin{Lem}
A necessary and sufficient condition for a monomial \ac{SC-LDPC-CC} with $w=c=3$ to have girth $g\geq10$ is that its symbolic parity-check matrix is free of repeated differences of differences.
\label{lem:diffsofdiffs}
\end{Lem}

\begin{IEEEproof}
For $w=3$ the following equation holds
\begin{equation}
\begin{split}
&\Delta p_{j}^{(i_1\leftrightarrow i_2)}+\Delta p_{j}^{(i_2\leftrightarrow i_3)}=\Delta p_{j}^{(i_1\leftrightarrow i_3)} \\ & \forall j, \forall i_3, \forall i_2 \neq i_3, \forall i_1 \neq i_2, i_3.
\end{split}
\label{eq:triwe}
\end{equation}
A generic cycle having length $6$ can be described as
\begin{equation}
\begin{split}
&\Delta p_{j_1}^{(i_1\leftrightarrow i_2)}+\Delta p_{j_2}^{(i_2\leftrightarrow i_3)}=\Delta p_{j_3}^{(i_1\leftrightarrow i_3)} \\ & \forall i_3, \forall i_2 \neq i_3, \forall i_1 \neq i_2, i_3, \forall j_3, \forall j_2 \neq j_3, \forall j_1 \neq j_2, j_3.
\end{split}
\label{eq:sixcycle}
\end{equation}
Substituting (\ref{eq:triwe}) for $j=j_1$ in (\ref{eq:sixcycle}) we obtain the alternative form
\begin{equation}
\Delta p_{j_1}^{(i_1\leftrightarrow i_3)} - \Delta p_{j_3}^{(i_1\leftrightarrow i_3)} = \Delta p_{j_1}^{(i_2\leftrightarrow i_3)} - \Delta p_{j_2}^{(i_2\leftrightarrow i_3)}.
\label{eq:sixsa}
\end{equation}
On the other hand, a cycle having length $8$ can be characterized, by definition, as
\begin{equation}
\Delta p_{j_1}^{(i_1\leftrightarrow i_2)} - \Delta p_{j_2}^{(i_1\leftrightarrow i_2)} = \Delta p_{j_3}^{(i_1\leftrightarrow i_3)} - \Delta p_{j_4}^{(i_1\leftrightarrow i_3)}.
\label{eq:eitsa}
\end{equation}
From (\ref{eq:sixsa}) and (\ref{eq:eitsa}) we see that, avoiding coincident differences of differences, we avoid cycles with both lengths $6$ and $8$. It is also true that (\ref{eq:sixsa}) and (\ref{eq:eitsa}) cover all the cases of equalities that can be formed from (\ref{eq:gendiffs}). Therefore, the converse holds as well.
\end{IEEEproof}
\begin{Cor}
For $w=c=3$, in order to avoid coincident differences of differences and achieve $g\geq 10$ we need
\begin{equation}
2 m_h \ge 3 \binom{a}{2}.
\label{eq:monwc3g10}
\end{equation}
\label{cor:cordiffs}
\end{Cor}
\begin{IEEEproof}
The absolute value of any difference of differences lies in the range $\left[1, 2,\ldots, 2m_h\right]$. For $w=c=3$, The total number of differences of differences is $3\binom{a}{2}$. By imposing that the cardinality of the set of possible values is greater than or equal to the total number of possibilities, we obtain (\ref{eq:monwc3g10}).
\end{IEEEproof}

The same reasoning can be extended to the case of $w = c = 2$ but, as we have seen before, in such a case cycles with length $10$ do not exist.
Therefore the following corollary holds.
\begin{Cor}
For $w=c=2$, in order to avoid equal differences of differences and achieve $g\geq 12$ we need
\begin{equation}
2 m_h \ge \binom{a}{2}.
\label{eq:g12monw2}
\end{equation}
\end{Cor}
\begin{IEEEproof}
Same as Corollary \ref{cor:cordiffs}.
\end{IEEEproof}

\subsubsection{Comparison with \ac{QC-LDPC} code matrices}

As shown in the previous sections, we can describe an \ac{SC-LDPC-CC} with a symbolic parity-check matrix $H(x)$.
Such a representation can also be used for \ac{QC-LDPC} block codes, the difference being in the procedure that converts $H(x)$ into the binary parity-check matrix $\HH$.
For \ac{QC-LDPC} block codes, the latter involves expansion of each $H(x)$ entry into a binary circulant block, while for \ac{SC-LDPC-CCs} (\ref{eq:bintopol}) is used.
Nevertheless, we can consider $H(x)$ matrices designed for \ac{QC-LDPC} block codes and use them to obtain \ac{SC-LDPC-CCs}. However, as we show next, this approach does not produce optimal results from the syndrome former constraint length standpoint.

Fossorier in \cite{Fossorier2004} provides two bounds for the symbolic matrix of \ac{QC-LDPC} codes. Both of them relate the minimum size of the circulant permutation blocks forming the code parity-check matrix to the code girth. 
The symbolic matrices considered in \cite{Fossorier2004} define Type-$1$ codes and contain all one entries in their first row and column, i.e.,

\begin{equation}
H(x)=\left[\begin{array}{llll}
1 & 1 & \ldots & 1\\
1 & x^{p_{1,1}} & \ldots & x^{p_{1,a-1}}\\
\vdots & \vdots & \ddots & \vdots\\
1 & x^{p_{c-1,1}} & \ldots & x^{p_{c-1,a-1}}\end{array}\right].
\label{eq:HxFos}
\end{equation}

In (\ref{eq:HxFos}), $p_{i,j}$ represents the cyclic shift to the right of the elements in each row of an identity matrix, and defines the circulant permutation matrix at position $\left( i,j \right)$. This representation has also been used in \cite{Tasdighi2016} and \cite{BocharovaHugJohannessonEtAl2012}.

Theorem $2.2$ in \cite{Fossorier2004} states that a necessary condition to have $g \geq 6$ is $p_{i,j_1} \neq p_{i, j_2}$ and $p_{i_1, j} \neq p_{i_2, j}$.
This theorem is also valid for \ac{SC-LDPC-CCs} and we have used a similar approach to demonstrate Lemma \ref{lem:lemmaonmonsw3}.
A corollary of this theorem claims that a necessary condition to have $g \geq 6$ in the Tanner
graph representation of a $(J, L)$-regular \ac{QC-LDPC} code is $p\geq L$ if $L$ is odd and $p>L$ if $L$ is even ($p$ is the size of the circulant permutation matrices).
The bounds on symbolic matrices resulting from this analysis can be extended to \ac{SC-LDPC-CCs}. In fact, for a \ac{QC-LDPC} code the maximum exponent appearing in $H(x)$ is equal to $p - 1$.
Therefore, if we use the symbolic matrix to define an \ac{SC-LDPC-CC}, the condition for $g=6$ becomes $\max\{{m_h}\} = p - 1$. Hence we find that $m_h\geq a-1$ if $a$ is odd and $m_h>a-1$ if $a$ is even.
By comparing these conditions with (\ref{eq:LhBoundCycles4w3monomial}), we see that having removed the border of ones in $H(x)$ has allowed us to obtain values of $m_h$ which are about half those needed in the presence of this constraint as in \cite{Fossorier2004}.
Examples of matrices with the smallest achievable $m_h$, $w=c=3$ and $g=6$ with and without the all one border of ones are as follows

\begin{equation}
H_1(x) =\left[\begin{array}{llll}
1 & 1 & 1 & 1\\
1 & x & x^2 & x^4\\
1 & x^3 & x & x^2\end{array}\right],
\label{eq:Hxcor}
\end{equation}

\begin{equation}
H_2(x) =\left[\begin{array}{llll}
1 & x^2 & 1 & 1\\
1 & 1 & x & x^2\\
x^2 & x & x & 1\end{array}\right].
\label{eq:Hxnocor}
\end{equation}
Indeed, the maximum exponent appearing in $H_1(x)$ is twice that in $H_2(x)$.
According to Theorem $2.4$ in \cite{Fossorier2004}, for $J \geq 3$ and $L \geq 3$, a necessary condition to have $g \geq 8$ is $p_{j_1,l_1} \neq p_{j_2,l_2}$ for $0<j_1<j_2$ and $0<l_1<l_2$.
A corollary of this theorem states that a necessary condition is $p > (J - 1)(L - 1)$. It is shown in \cite{Tasdighi2016} that this bound is usually not tight when $J\geq 4$.

Extending these results to \ac{SC-LDPC-CCs}, we can calculate a necessary condition to avoid cycles with length $4$ and $6$ in codes with the all one border, thus obtaining $m_h\geq(a-1)(c-1)$.
Also in this case, the all one border causes an increase in $m_h$, as it is evident from the following example

\begin{equation}
H_1(x)=\left[\begin{array}{llll}
1 & 1 & 1 & 1\\
1 & x^6 & x^2 & x^5\\
1 & x^3 & x^4 & x\end{array}\right],
\label{eq:Hxcorg8}
\end{equation}

\begin{equation}
H_2(x)=\left[\begin{array}{llll}
1 & x^3 & 1 & x^3\\
x^3 & x^2 & x^2 & 1\\
x^2 & 1 & x^3 & x^2\end{array}\right].
\label{eq:Hxnocorg8}
\end{equation}

An intermediate step for the generalization of the codes in \cite{Fossorier2004} is the removal of the constraint of having the first column filled with ones, that is,
\begin{equation}
H(x)=\left[\begin{array}{llll}
1 & 1 & \ldots & 1\\
x^{p_{1,0}} & x^{p_{1,1}} & \ldots & x^{p_{1,a-1}}\\
\vdots & \vdots & \ddots & \vdots\\
x^{p_{c-1,0}} & x^{p_{c-1,1}} & \ldots & x^{p_{c-1,a-1}}\end{array}\right].
\label{eq:HxSupBo}
\end{equation}
As we show next, this modification leads to a reduction in the lower bounds on $m_h$. Codes having symbolic matrix as in (\ref{eq:HxSupBo}) are referred to as Type-1 constrained (Type-1c) codes.

\begin{Lem}
A necessary condition for a Type-1c \ac{SC-LDPC-CC} with $w=c=3$ to achieve $g\ge 6$ is $m_h \geq a - 1$, $\forall a$.
\label{Lem: FossMOd1}
\end{Lem}
\begin{IEEEproof}
Similar to the proof of Corollary 2.2 in \cite{Fossorier2004}.
\end{IEEEproof}

\begin{Lem}
A necessary condition for a Type-1c \ac{SC-LDPC-CC} with $w=c=3$ to achieve $g\ge 8$ is $m_h \geq \frac{a(c-1)}{2}-1$.
\label{Lem: FossMOd2}
\end{Lem}
\begin{IEEEproof}
A cycle with length $6$ follows from equalities having the following form: $p_{j_1,l_1}+p_{j_2,l_2}=p_{j_1,l_2}+p_{j_2,l_3}$. If there exists some $p^*_{j_1,l_1}=p^*_{j_2,l_1}$, then to avoid length-$6$ cycles there must not be any $p^d_{j_{1},l_{1}}=p^d_{j_{2},l_{2}}$ in the remaining entries of the matrix.  Otherwise, either a cycle with length $6$ resulting from $p^d_{j_{1},l_{1}}+p^*_{j_1,l_1}=p^d_{j_{2},l_{2}}+p^*_{j_2,l_1}$ or a cycle with length $4$ resulting from $p^d_{j_{1},l_{1}}-p^d_{j_2,l_{2}}=p^*_{j_1,l_1}-p^*_{j_2,l_1}$ would appear. In this case, we would obtain $m_h\geq(a-1)(c-1)$, showing that the case studied in \cite{Fossorier2004} is only a particular case of the above construction. If $p_{j_1,l_1}=p_{j_2,l_1}$ is never verified, we can fill the $(c-1)a$ possible values with $\frac{(c-1)a}{2}$ couples of exponents (also $0$ is admitted), leading to $m_h \geq \frac{a(c-1)}{2}-1$.
\end{IEEEproof}

The lower bounds provided by Lemma \ref{Lem: FossMOd1} and Lemma \ref{Lem: FossMOd2} remain worse than those obtained with the general structure of $H(x)$, thus confirming that removal of the constraints in (\ref{eq:HxFos}) is preferable in view of minimizing $m_h$.

\subsection{Type-$z$ codes}

In order to meet the condition $g \geq 6$, we must ensure that cycles with length $4$ do not exist.
Such short cycles occur when, for some $i_1,j_1,i_2,j_2$, $j_1 \neq j_2$,
\begin{equation}
\delta_{i_1,j_1} = \delta_{i_2,j_2} \quad \textrm{and} \quad s_{l_1} = s_{l_2},
\label{eq:ConditionCycles4}
\end{equation}
where $s_{l_1}$ and $s_{l_2}$ are the starting levels of $\delta_{i_1, j_1}$ and $\delta_{i_2, j_2}$, respectively.
So, in order to avoid cycles with length $4$, there must not be any two equal differences starting from the same level.
We observe that such two differences may even be in the same column of $\HH_s^T$.

\begin{Lem}
A Type-$z$ code with $\HH_s^T$ having $w=2$, free of length-$4$ cycles, has
\begin{equation}
a \le  \sum_{i=0}^{c-1} (L_h - i -1) = c L_h - \binom{c+1}{2},
\label{eq:aupperboundCycles4}
\end{equation}
that is 
\[
L_h \ge \Bigg\lceil \frac{a+\binom{c+1}{2}}{c} \Bigg\rceil.
\]
Considering that by construction it must be $L_h > c$, we obtain
\begin{equation}
L_h \ge \max \Bigg\{c+1, \Bigg\lceil \frac{a+\binom{c+1}{2}}{c} \Bigg\rceil \Bigg\}.
\label{eq:LhBoundw2Cycles4}
\end{equation}
\label{lem:LhboundCycles4}
\end{Lem}

\begin{IEEEproof}
For column weight $w=2$, each column of $\HH_s^T$ only contains one difference $\delta_{i,j}$ and each difference can be used up to $c$ times without incurring cycles with length $4$ (by using all the possible $c$ starting levels).
For a given $L_h$, the number of possible differences starting from level $l$ is $L_h-l-1$.
Since the differences corresponding to any two of the $a$ columns of $\HH_s^T$ must be different
in value and/or starting level, summing all the contributions we obtain (\ref{eq:aupperboundCycles4}), from which (\ref{eq:LhBoundw2Cycles4}) is eventually derived.
\end{IEEEproof}

We can extend \eqref{eq:aupperboundCycles4} to the case of a regular $\HH_s^T$ with $w > 2$
by considering that, in such a case, each column of $\HH_s^T$ provides $\binom{w}{2}$ differences that
must meet condition \eqref{eq:ConditionCycles4}.
Hence, \eqref{eq:aupperboundCycles4} becomes
\[
a\binom{w}{2} \le c L_h - \binom{c+1}{2},
\]
while \eqref{eq:LhBoundw2Cycles4} becomes
\begin{equation}
L_h \ge \max \Bigg\{c+1, \Bigg\lceil \frac{a \binom{w}{2}+\binom{c+1}{2}}{c} \Bigg\rceil \Bigg\}.
\label{eq:LhBoundCycles4Regular}
\end{equation}

When we have an irregular $\HH_s^T$ having different columns weights $w_i$, $i=0,1,2,\ldots, a-1$, each column of $\HH_s^T$ 
corresponds to $\binom{w_i}{2}$ differences. Therefore \eqref{eq:LhBoundCycles4Regular} becomes
\begin{equation}
L_h \ge \max \Bigg\{c+1, \Bigg\lceil \frac{ \sum_{i=0}^{a-1}{\binom{w_i}{2}}+{\binom{c+1}{2}}}{c} \Bigg\rceil \Bigg\}.
\label{eq:LhBoundCycles4Irregular}
\end{equation}

In order to find the conditions which permit us to have $g \geq 8$, let us first consider the case with $c=1$ and $\HH_s^T$ with column weight $w=2$.
Since the sum of two odd integers is even, the following proposition easily follows.
\begin{Pro}
For $c=1$ and $w=2$, if all the $\delta_{i,j}$ are different and odd, then $g \geq 8$.
\label{luno}
\end{Pro}

From Proposition \ref{luno} it follows that, if we wish to minimize $L_h$, we can choose the values of $\delta_{i,j}$ equal to $\left\{1,3,5,\ldots,2a-1\right\}$ and the code will be free of cycles with length smaller than $8$.

\begin{Lem}
For $c=1$ and $w=2$, cycles having length smaller than $8$ can be avoided if and only if
\begin{equation}
L_h \ge 2a.
\label{eq:LhBoundc1w2Cycles8}
\end{equation}
\label{lem:LhBoundc1w2Cycles8}
\end{Lem}
\begin{IEEEproof}
From Proposition \ref{luno} 
we see that the maximum value of a difference that is needed to avoid cycles with length smaller than $8$ is $2a-1$.
Therefore, we have $L_h \ge 1+2a-1=2a$.
In order to prove the converse, let us consider that from the set $\left[1, 2,\ldots, y\right]$ we can select at most $\frac{y}{2}$ values which may be summed pairwise resulting in other values in the same set.
If we choose the values of the differences from the set $\left[1, 2, \ldots, 2a - 2\right]$, we only have $a - 1$ values which may be summed pairwise resulting in other values in the same set.
Therefore, we can only allocate $a - 1$ differences without introducing cycles with length $6$, which is not sufficient to cover all the $a$ rows of $\HH_s$.
So, $[1, 2, \ldots, 2a-1]$ is the smallest set of difference values able to avoid cycles with length $4$ and $6$, from which \eqref{eq:LhBoundc1w2Cycles8} follows.
\end{IEEEproof}

Equation \eqref{eq:LhBoundc1w2Cycles8} can be extended to the case $c > 1$ by considering that, in such a case, each difference value can be repeated up to $c$ times (by exploiting all the $c$
available values as starting levels). Therefore, for $w = 2$ and $c > 1$ we have
\begin{equation}
L_h \ge \max \left\{c+1, \frac{2a}{c} \right\}.
\label{eq:LhBoundw2Cycles8}
\end{equation}

Let us consider larger values of $w$, i.e., $w \ge 3$.
For $c=1$, each column of $\HH_s^T$ has one or more cycles with length $6$, since at least $3$ symbols $1$ are at the same level (as described in Section \ref{sec:LocalCycles}).

Instead, for $w \ge 3$ and $c>1$ we can follow the same approach used for the case with $g = 6$. This way, we obtain that to have $g \geq 8$ we need
\begin{equation}
L_h \ge \max \left\{c+1, \frac{2a \binom{w}{2}}{c} \right\}.
\label{eq:LhBoundCycles8}
\end{equation}
 
Finally, when the columns of $\HH_s^T$ are irregular with weights $w_i$, $i=0,1,2,\ldots, a-1$, as done for the case with $g = 6$, we can consider that each row of $\HH_s$ corresponds to $\binom{w_i}{2}$
differences and \eqref{eq:LhBoundCycles8} becomes
\begin{equation}
L_h \ge \max \Bigg\{c+1, \Bigg\lceil \frac{2 \sum_{i=0}^{a-1}{\binom{w_i}{2}}}{c} \Bigg\rceil \Bigg\}.
\label{eq:LhBoundCycles8Irreg}
\end{equation}

\section{Design Methods \label{sec:Methods}}

In this section we introduce new methods for the design of Type-1, Type-2 and Type-3 \ac{SC-LDPC-CCs} with constraint length approaching the bounds described in Section \ref{sec:MinConstLength}. 
These design approaches have general validity. However, for the sake of simplicity we focus on the case $w=c=3$. 

\subsection{Type-1 codes \label{subsec:ty1meths}}

As we have already demonstrated in Lemma \ref{lem:lemmaonmonsw3}, monomial codes having $g \geq 6$ satisfy (\ref{eq:LhBoundCycles4w3monomial}).
For any integer value $k$, a Type-1 code with $a = 2k +1$ able to achieve such a bound is defined by the symbolic matrix as in (\ref{eq:matrminmhg6even}),
\begin{figure*}[!t]
\begin{equation}
\resizebox{.9 \textwidth}{!}
{
$H_e(x,k)=\left[\begin{array}{cccccccccccc}
\arraycolsep=1.4pt\def\arraystretch{5pt}
x^k & x^k & x^k & \ldots & x^k & x^k & 1 & 1 & 1 & \ldots & 1 & 1 \\
1 & x & x^2 & \ldots & x^{k-1} & x^k & x & x^2 & x^3 & \ldots & x^{k-1} & x^k \\
x^k & x^{k-1} & x^{k-2} & \ldots & x & 1 & x^k & x^{k-1} & x^{k-2} & \ldots & x^2 & x\\
\end{array}\right]$
\label{eq:matrminmhg6even}
}
\end{equation}
\end{figure*}
when $k$ is even and by (\ref{eq:matrminmhg6odd})
\begin{figure*}[!t]
\begin{equation}
\resizebox{.9 \textwidth}{!}
{
$H_o(x,k)=\left[\begin{array}{cccccccccccc}
\arraycolsep=1.4pt\def\arraystretch{5pt}
x^k & x^k & x^k & \ldots & x^k & x^k & 1 & 1 & 1 & \ldots & 1 & 1 \\
1 & x & x^2 & \ldots & x^{k-2} & x^{k-1} & 1 & x & x^2 & \ldots & x^{k-1} & x^k \\
x^{k-1} & x^{k-2} & x^{k-3} & \ldots & x & 1 & x^k & x^{k-1} & x^{k-2} & \ldots & x & 1\\
\end{array}\right]$
}
\label{eq:matrminmhg6odd}
\end{equation}
\end{figure*}
when $k$ is odd.
Starting from (\ref{eq:matrminmhg6even}) and (\ref{eq:matrminmhg6odd}), the removal of one column permits us to also cover the case of $a = 2k$.

\subsection{Type-2 codes \label{subsec:ty2meths}}

In the case of Type-2 codes with $w=c=3$, we can jointly use monomial and binomial entries in the same column of the symbolic parity-check matrix. The bound given by (\ref{eq:LhBoundCycles4Regular}), expressed in terms of $m_h$, becomes
\begin{equation}
m_h \ge  \Big\lceil \frac{a-1}{3} \Big\rceil.
\label{eq:limitw3c3onmh}
\end{equation}

Let $k$ be an integer and $H_e^1(x,k)$, $H_e^2(x,k)$ and $H_e^3(x,k)$ be the $3\times k$ symbolic matrices as in \Cref{eq:bineven1,eq:bineven2,eq:bineven3}
and let $H_o^1(x,k)$, $H_o^2(x,k)$ and $H_o^3(x,k)$ be the $3\times k$ symbolic matrices as in \Cref{eq:binodd1,eq:binodd2,eq:binodd3}.
\begin{figure*}[!t]
\begin{equation}
\resizebox{.9 \textwidth}{!}
{$H_e^1(x,k)=\left[\begin{array}{cccccccc}
0 & 0 & \ldots & 0 & 0 & 0 &  \ldots & 0\\
x+x^k & x^2+x^{k-1} & \ldots & x^{\frac{k}{2}}+x^{\frac{k+2}{2}} & 1 & 1 & \ldots & 1  \\
1 & 1 & \ldots & 1 & 1+x^{k} & x+x^{k-1} & \ldots & x^{\frac{k-2}{2}}+x^{\frac{k+2}{2}}
\end{array}\right]$}
\label{eq:bineven1}
\end{equation}
\end{figure*}
\begin{figure*}[!t]
\begin{equation}
\resizebox{.9 \textwidth}{!}
{$H_e^2(x,k)=\left[\begin{array}{cccccccc}
1 & 1 & \ldots & 1 & 1+x^{k} & x+x^{k-1} & \ldots & x^{\frac{k-2}{2}}+x^{\frac{k+2}{2}} \\
0 & 0 & \ldots & 0 & 0 & 0 &  \ldots & 0\\
x+x^k & x^2+x^{k-1} & \ldots & x^{\frac{k}{2}}+x^{\frac{k+2}{2}} & 1 & 1 & \ldots & 1
\end{array}\right]$}
\label{eq:bineven2}
\end{equation}
\end{figure*}
\begin{figure*}[!t]
\begin{equation}
\resizebox{.9 \textwidth}{!}
{$H_e^3(x,k)=\left[\begin{array}{cccccccc}
x+x^k & x^2+x^{k-1} & \ldots & x^{\frac{k}{2}}+x^{\frac{k+2}{2}} & 1 & 1 & \ldots & 1  \\
1 & 1 & \ldots & 1 & 1+x^{k} & x+x^{k-1} & \ldots & x^{\frac{k-2}{2}}+x^{\frac{k+2}{2}} \\
0 & 0 & \ldots & 0 & 0 & 0 &  \ldots & 0
\end{array}\right]$}
\label{eq:bineven3}
\end{equation}
\end{figure*}
\begin{figure*}[!t]
\begin{equation}
\resizebox{.9 \textwidth}{!}
{$H_o^1(x,k)=\left[\begin{array}{cccccccc}
0 & 0 & \ldots & 0 & 0 & 0 &  \ldots & 0\\
x+x^k & x^2+x^{k-1} & \ldots & x^{\frac{k-1}{2}}+x^{\frac{k+3}{2}} & 1 & 1 & \ldots & 1  \\
1 & 1 & \ldots & 1 & 1+x^{k} & x+x^{k-1} & \ldots & x^{\frac{k-1}{2}}+x^{\frac{k+1}{2}}
\end{array}\right]$}
\label{eq:binodd1}
\end{equation}
\end{figure*}
\begin{figure*}[!t]
\begin{equation}
\resizebox{.9 \textwidth}{!}
{$H_o^2(x,k)=\left[\begin{array}{cccccccc}
1 & 1 & \ldots & 1 & 1+x^{k} & x+x^{k-1} & \ldots & x^{\frac{k-1}{2}}+x^{\frac{k+1}{2}} \\
0 & 0 & \ldots & 0 & 0 & 0 &  \ldots & 0\\
x+x^k & x^2+x^{k-1} & \ldots & x^{\frac{k-1}{2}}+x^{\frac{k+3}{2}} & 1 & 1 & \ldots & 1
\end{array}\right]$}
\label{eq:binodd2}
\end{equation}
\end{figure*}
\begin{figure*}[!t]
\begin{equation}
\resizebox{.9 \textwidth}{!}
{$H_o^3(x,k)=\left[\begin{array}{cccccccc}
x+x^k & x^2+x^{k-1} & \ldots & x^{\frac{k-1}{2}}+x^{\frac{k+3}{2}} & 1 & 1 & \ldots & 1  \\
1 & 1 & \ldots & 1 & 1+x^{k} & x+x^{k-1} & \ldots & x^{\frac{k-1}{2}}+x^{\frac{k+1}{2}} \\
0 & 0 & \ldots & 0 & 0 & 0 &  \ldots & 0
\end{array}\right]$}
\label{eq:binodd3}
\end{equation}
\end{figure*}

It is easy to verify that each one of the matrices in (\ref{eq:bineven1}) to (\ref{eq:binodd3}) results in a code with girth $6$.
For even and odd values of $k$, codes with $a=3k$ are defined by, respectively,
\begin{equation}
H_e(x,k)=\left[\begin{array}{ccc}
H_e^1(x,k)\;\;H_e^2(x,k)\;\;H_e^3(x,k)
\end{array} \right]
\label{eq:binevenminmh}
\end{equation}
\begin{equation}
H_o(x,k)=\left[\begin{array}{ccc}
H_o^1(x,k)\;\;H_o^2(x,k)\;\;H_o^3(x,k)
\end{array} \right].
\label{eq:binoddminmh}
\end{equation}
Also in this case, it is possible to remove one or more columns to obtain the required values of $a$. In fact, both $H_e(x,k)$ and $H_o(x,k)$ define codes with girth $6$. Any reduction of the girth with respect to their component matrices is avoided owing to the order of the rows of $H^i_e(x,k)$ and $H^i_o(x,k)$ ($i=1,2,3$), which ensures that the supports of any two columns belonging to two different component matrices do not overlap in more than one position.

\begin{Exa}\label{Ex:1}
Let $w=c=3$ and $a=3k=12$ (i.e., $k=4$). Based on (\ref{eq:limitw3c3onmh}), $m_{h}\geq \Big\lceil\dfrac{12-1}{3}\Big\rceil=4$. By concatenating the three matrices in (\ref{eq:trismatrices}), each of size $3\times 4$, it is possible to construct a syndrome former matrix $H_e(x,4)$ with $m_{h}=4$.
\begin{figure*}[!t]
\begingroup\fontsize{9.5pt}{7pt}
\begin{equation}
\begin{split}
H_e^1(x,k)=&\left[\begin{array}{cccc}
0 & 0 & 0 & 0\\
x+x^4 & x^2+x^{3} &1 & 1 \\
1 & 1 &1+x^4 & x+x^{3}
\end{array}\right],\;\;\; 
H_e^2(x,k)=\left[\begin{array}{cccc}
1 & 1 &1+x^4 & x+x^{3}\\
0 & 0 & 0 & 0 \\
x+x^4 & x^2+x^{3} &1 & 1
\end{array}\right],
\\&H_e^3(x,k)=\left[\begin{array}{cccc}
 x+x^4 & x^2+x^{3} &1 & 1\\
1 & 1 &1+x^4 & x+x^{3}\\
0 & 0 & 0 & 0
\end{array}\right]
\end{split}
\label{eq:trismatrices}
 \end{equation}
\endgroup
\end{figure*}
\end{Exa}

\subsection{Type-3 codes \label{subsec:ty3meths}}

Finally, let us consider the use of trinomial entries in the columns of the symbolic parity-check matrices. Since $w=3$, we cannot use more than one trinomial entry per column but it is possible to use every trinomial entry three times, one per each row of the symbolic matrix.
For different values of $a$, some groups of trinomials with the smallest possible degrees that permit us to avoid cycles with length $4$ are shown in (\ref{eq:Trins}).

\begin{figure*}[!t]
\begin{equation}
\resizebox{.9 \textwidth}{!}
{
$\begin{cases}
1+x+x^4 \quad 1+x^2+x^7  \quad a=6\\
1+x+x^5 \quad 1+x^6+x^8 \quad 1+x^3+x^{10} \quad a=9 \\
1+x^2+x^9 \quad 1+x^3+x^8 \quad 1+x^4+x^{10} \quad 1+x+x^{12} \quad a=12 \\
1+x^{11}+x^{12} \quad 1+x^{13}+x^{15} \quad 1+x^7+x^{10} \quad 1+x^5+x^9 \quad 1+x^8+x^{14} \quad a=15\\
1+x^6+x^7 \quad 1+x^{13}+x^{15} \quad 1+x^{14}+x^{17} \quad 1+x^8+x^{12} \quad 1+x^{11}+x^{16} \quad 1+x^{10}+x^{19} \quad a=18 \\
\ldots
\end{cases}$
}
\label{eq:Trins}
\end{equation}
\end{figure*}

To see why these trinomials are those with smallest degrees, we refer the interested reader to \cite{HPark2}, where the authors exploit a combinatorial notion known as Perfect Difference Families (PDF). They show that Type-$3$ \ac{QC-LDPC} block codes with $w=3,\;c=1,\;a=3k$ and girth $6$ can be constructed by using the entries of each block of a $(v,3,1)$ PDF to define a trinomial in the syndrome former matrix. In fact, if our target syndrome former matrix is only formed by trinomials, PDF's give us the smallest trinomial degrees. Those reported in (\ref{eq:Trins}) and many other groups of trinomials can be found starting from \cite[Table I]{Tasdighi2016b}.

We also notice that we can combine trinomial and monomial entries to design codes with $g \geq 6 $. In fact, trinomial entries introduce ``horizontal'' separations whereas monomial entries introduce ``vertical'' separations. This is also the reason why, in general, trinomial and binomial entries cannot be combined in the configurations described in Section \ref{subsec:ty2meths}. 
We provide in (\ref{eq:oddsplit}) an example in which we concatenate two matrices having the form (\ref{eq:matrminmhg6odd}) and (\ref{eq:Trins}) to construct a syndrome former matrix with $w=c=3$, $a=21$ and girth $6$, achieving the smallest possible $m_{h}$ according to (\ref{eq:limitw3c3onmh}).
\begin{figure*}[!t]
\begingroup\small\begin{equation}
\begin{split}
H(x,7)_{odd}=&\left[\begin{array}{cccccc}
1+x+x^4 & 1+x^2+x^7  & 0 & 0 & 0 & 0 \\
 0 & 0 & 1+x+x^4 & 1+x^2+x^7 & 0 & 0 \\
 0 & 0 & 0 & 0 & 1+x+x^4 & 1+x^2+x^7
\end{array}\right| \ldots\\
&\left|\begin{array}{ccccccccccccccc}
 x^7 & x^7 & x^7 & x^7 & x^7 & x^7 & x^7 & 1 &1  &1  & 1 & 1   & 1 &1  &1\\
1 & x & x^2 & x^3 & x^4 & x^5 & x^6 & 1 & x & x^2 & x^3 &  x^4  & x^5 & x^6 &x^7\\
x^6 & x^5 & x^4 & x^3 & x^2 & x & 1 & x^7 & x^6 & x^5 & x^4 &   x^3 & x^2 & x & 1
\end{array} \right]
\end{split}
\label{eq:oddsplit}
\end{equation}
 \endgroup
\end{figure*}

\section{Code examples and numerical results\label{sec:Examples}}

\begin{figure*}[!t]
\centering
\subfloat[][]{\includegraphics[width=0.4\textwidth]{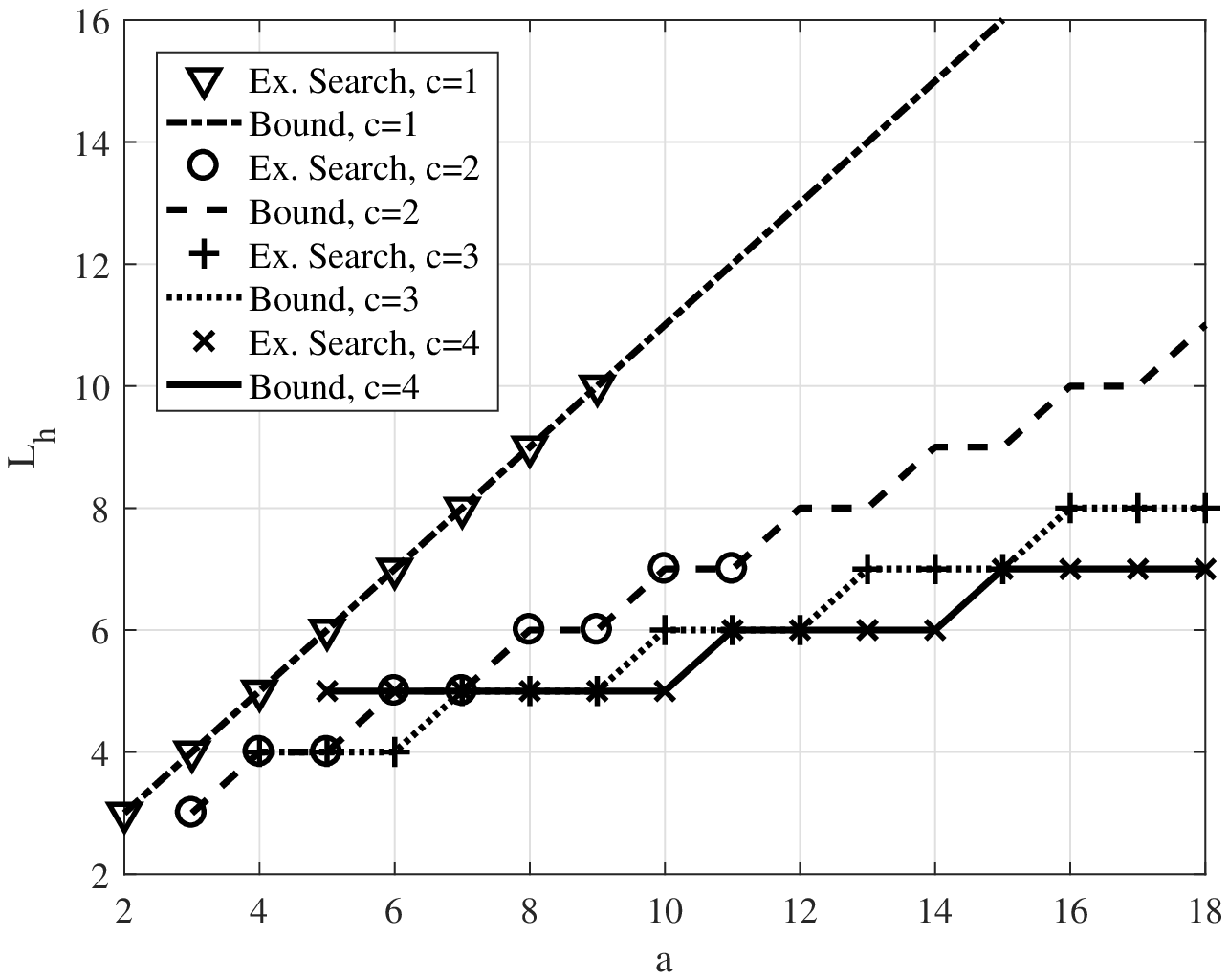}\label{fig:w2c1234g6}}
\subfloat[][]{\includegraphics[width=0.4\textwidth]{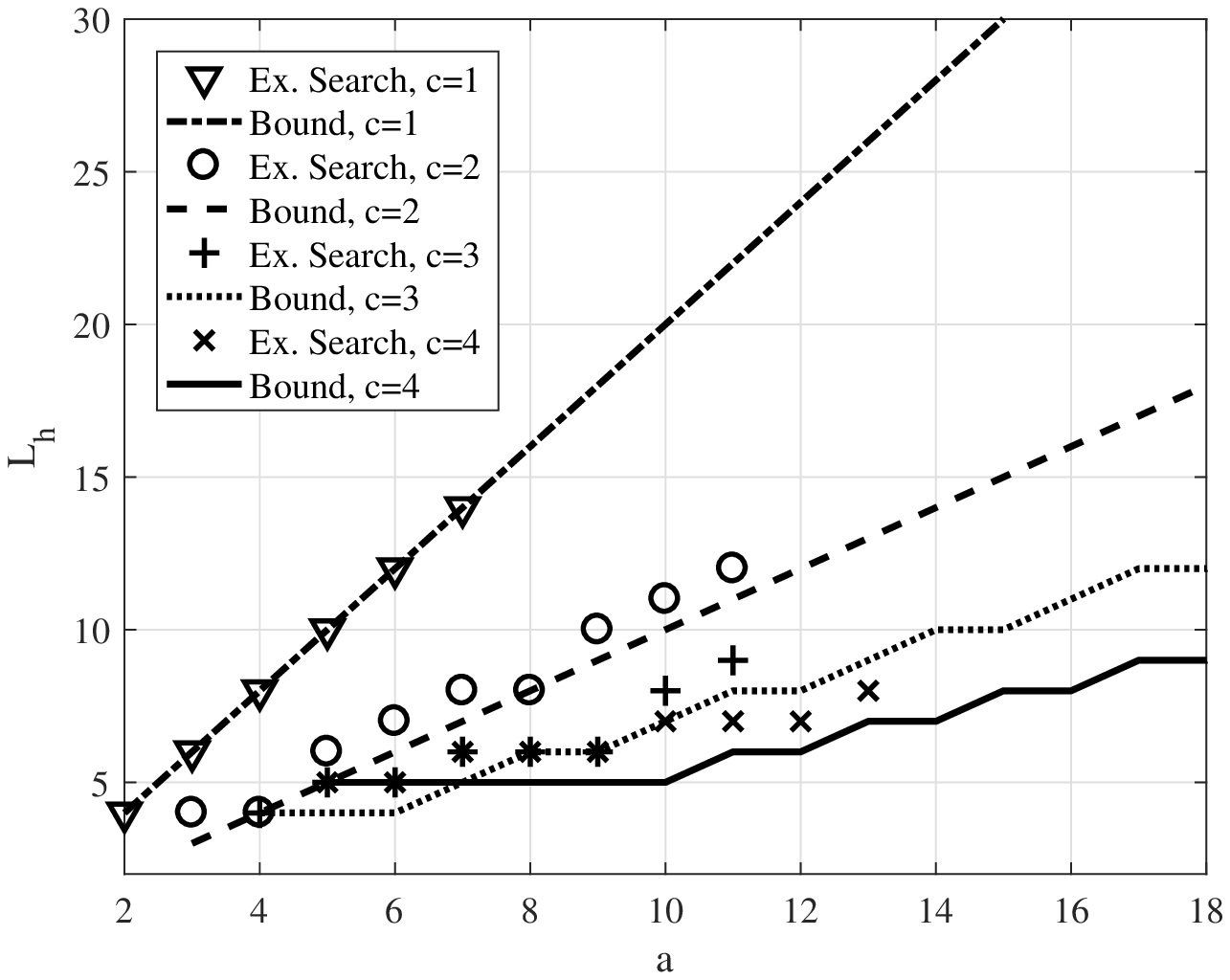}\label{fig:w2c1234g8}}
\vfill
\subfloat[][]{\includegraphics[width=0.4\textwidth]{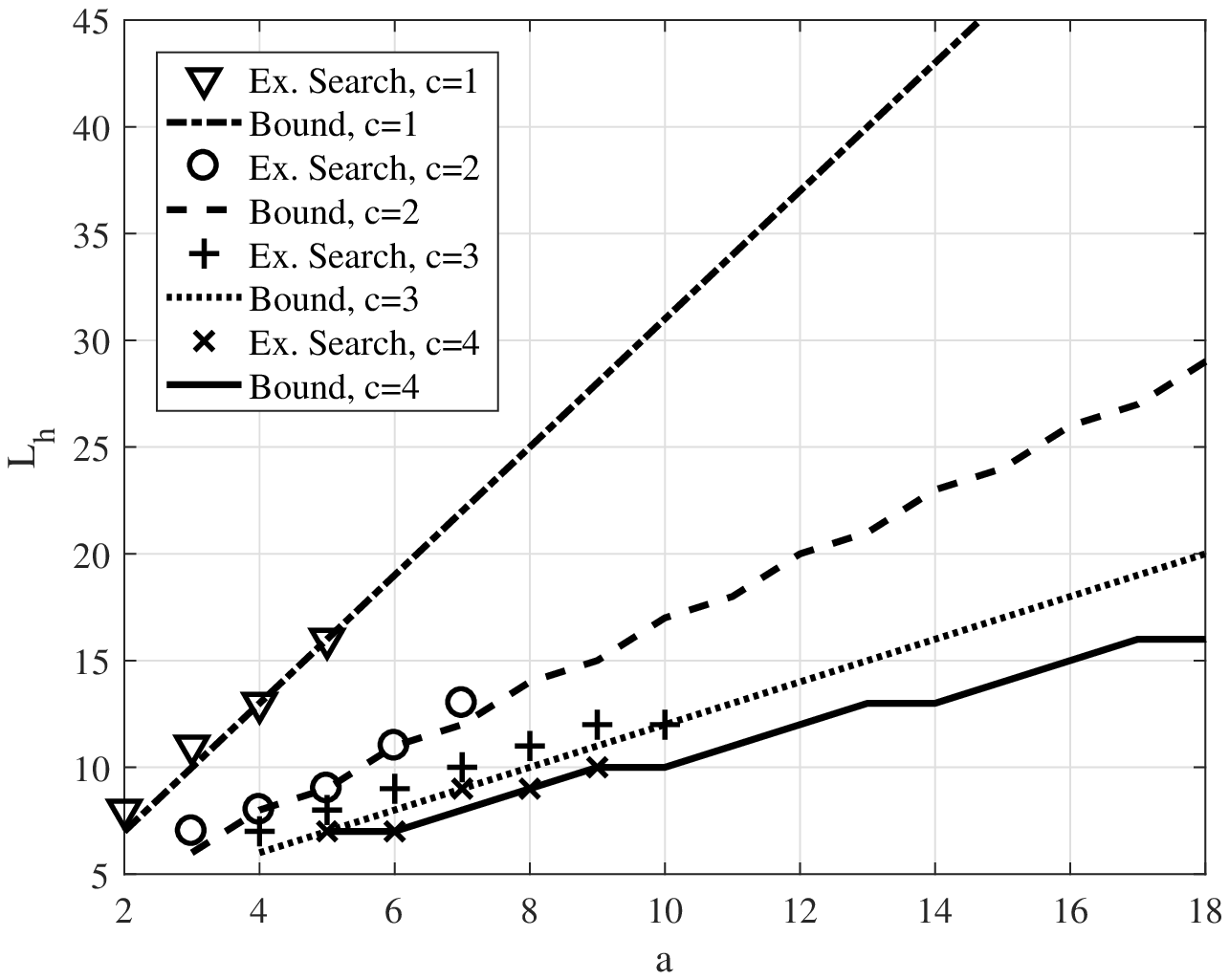}\label{fig:w3c1234g6}}
\subfloat[][]{\includegraphics[width=0.4\textwidth]{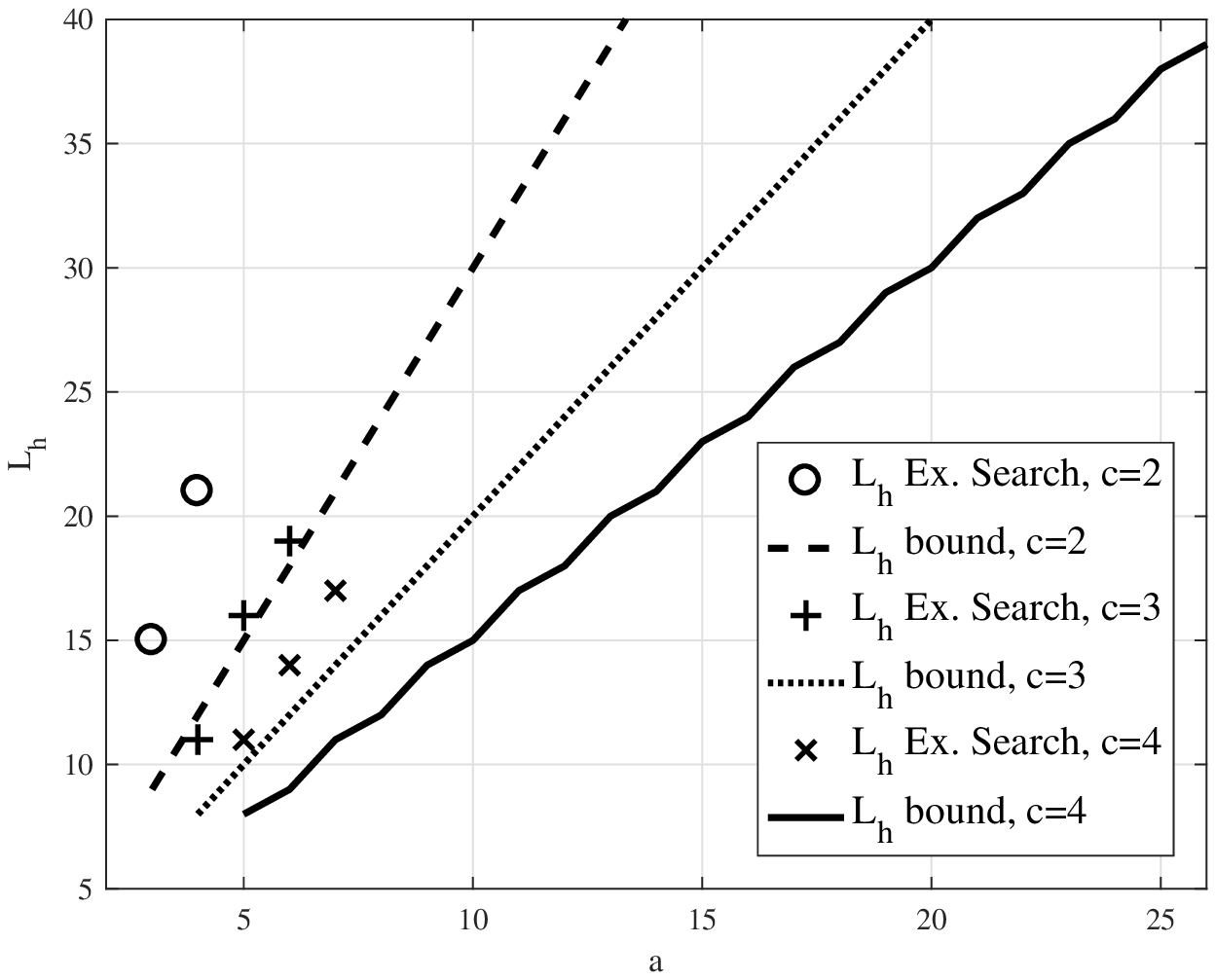}\label{fig:w3c234g8}}
 \caption{Bounds on $L_h$ and values found through exhaustive searches as a function of $a$ for some values of $c$ and: (a) $w=2$, $g=6$ (b) $w=2$, $g=8$ (c) $w=3$, $g=6$ (d) $w=3$, $g=8$.}
 \label{fig:aa}
\end{figure*}

In this section we provide some examples of code design and their comparison with analytical bounds. We also assess the codes performance through numerical simulations of coded transmissions.

\subsection{Code design based on exhaustive searches \label{subsec:Exha}}

In Fig. \ref{fig:aa} we report the bounds on $L_h$ obtained as described in Section \ref{sec:MinConstLength}
as a function of $a$, for some values of $w$, $g$ and $c$.
We compare these bounds with the results obtained through exhaustive searches over all the possible choices of $\HH_s^T$. The matching between the
bound and the values found through exhaustive searches is perfect for all the considered values of $c$ when $w=2$ and $g=6$, as shown in Fig. \ref{fig:w2c1234g6}. We note in Fig. \ref{fig:w2c1234g8} that the results of
exhaustive searches are well matched with the bounds also for the case with $w = 2$ and
$g = 8$. When $w > 2$, the
bound may not be achievable in practical terms. We observe in Fig. \ref{fig:w3c1234g6} that the deviations of the heuristic values from the theoretical curves are rather small when $g=6$. The gap to the bound increases for increasing values of $c$ when $g=8$, as can be noticed in Fig. \ref{fig:w3c234g8}.

\begin{figure}[t!]
\begin{centering}
\includegraphics[width=90mm,keepaspectratio]{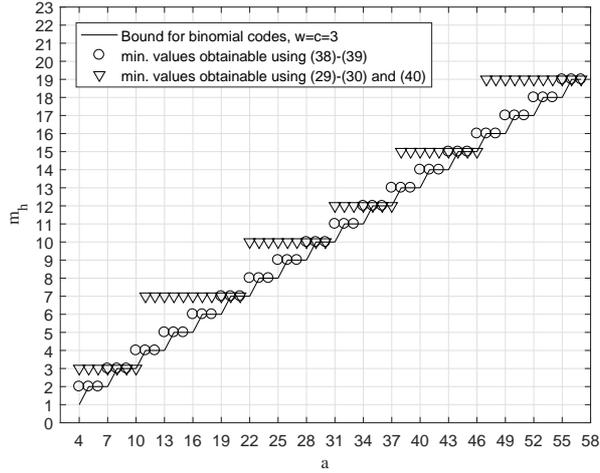}
\caption{Bounds on $m_h$ and values found through the design method described in Section \ref{sec:Methods} as a function of $a$, for $w=c=3$, $g=6$.}
\label{fig:w3c3meths}
\par\end{centering}
\end{figure}

\begin{table}[t]
\renewcommand{\arraystretch}{1}
\caption{Minimum values of $m_{h}$ for monomial codes with $w=c=3$ and $g=8$.}
\label{table:6mon}
\centering
\begin{tabular}{|@{\hspace{0.8mm}}c@{\hspace{0.8mm}}|@{\hspace{0.8mm}}c@{\hspace{0.8mm}}|@{\hspace{0.8mm}}c@{\hspace{0.8mm}}|@{\hspace{0.8mm}}c@{\hspace{0.8mm}}|@{\hspace{0.8mm}}c@{\hspace{0.8mm}}|}
\hline
 CODE RATE & $1/4$ & $2/5$ & $3/6$ & $4/7$ \\ [0.5ex] \hline
Exhaustive Search& $3$&	$5$	&$6$	& $8$ \\ 
\hline 

Bound (\ref{eq:lemg8w3mon}) & $2$&	$3$	&$4$	& $6$\\ 
\hline
\end{tabular}
\end{table}

In Fig. \ref{fig:w3c3meths} we consider $w = c = 3$, $g = 6$, and compare the values of $m_h$ achievable through the design methods described in Section \ref{sec:Methods} with the corresponding lower bounds.
In Table \ref{table:6mon} we instead provide the results of an exhaustive search of monomial codes with $w = c = 3$ and $g = 8$, and their comparison with the corresponding lower bound.

\subsection{Code design based on \ac{IP} \label{subsec:MonteCarlo}}

When the girth takes values $g \geq 8$, the values of $m_h$ or, equivalently, $L_h$ are so high that exhaustive searches become unfeasible even for small values of $a$ and $c$. Appendix B contains a brief description of the complexity of the searching algorithms.

\begin{table*}[!t]
\renewcommand{\arraystretch}{1.2}
\caption{Minimum values of $m_h$ for codes with $w=c=3,4,5,6$ and $g=8$ obtained through  the Min-Max algorithm, for several code rates.}
\label{table:Tabg8MinMax}
\centering
\begin{tabular}{|@{\hspace{0.3mm}}c@{\hspace{0.3mm}}|@{\hspace{0.3mm}}c@{\hspace{0.3mm}}|@{\hspace{0.3mm}}c@{\hspace{0.3mm}}|@{\hspace{0.3mm}}c@{\hspace{0.3mm}}|@{\hspace{0.3mm}}c@{\hspace{0.3mm}}|@{\hspace{0.3mm}}c@{\hspace{0.3mm}}|@{\hspace{0.3mm}}c@{\hspace{0.3mm}}|@{\hspace{0.3mm}}c@{\hspace{0.3mm}}|@{\hspace{0.3mm}}c@{\hspace{0.3mm}}|@{\hspace{0.3mm}}c@{\hspace{0.3mm}}|@{\hspace{0.3mm}}c@{\hspace{0.3mm}}|@{\hspace{0.3mm}}c@{\hspace{0.3mm}}|@{\hspace{0.3mm}}c@{\hspace{0.3mm}}|@{\hspace{0.3mm}}c@{\hspace{0.3mm}}|@{\hspace{0.3mm}}c@{\hspace{0.3mm}}|@{\hspace{0.3mm}}c@{\hspace{0.3mm}}|@{\hspace{0.3mm}}c@{\hspace{0.3mm}}|@{\hspace{0.3mm}}c@{\hspace{0.3mm}}|@{\hspace{0.3mm}}c@{\hspace{0.3mm}}|@{\hspace{0.3mm}}c@{\hspace{0.3mm}}|} 
\hline
\multirow{2}{*}{$w=c=3$}&Code rate& $1/4$ & $2/5$ & $3/6$ & $4/7$ & $5/8$ & $6/9$ & $7/10$ & $8/11$ & $9/12$ & $10/13$ &$11/14$ & $12/15$ & $13/16$ & $14/17$ & $15/18$ & $16/19$ & $17/20$ & $22/25$ 
\\ \cline{2-20}
&$m_h $&$3$&$6$&$7$&$9$&$10$&$12$&$15$ &$17$ &$21$ &$24$ &$26$ &$29$ &$32$ &$37$ &$39$ &$43$ &$48$ &$76$
  \\ \hline \hline
\multirow{2}{*}{$w=c=4$}&Code Rate &--& $1/5$ & $2/6$ & $3/7$ & $4/8$ & $5/9$ & $6/10$ & $7/11$ & $8/12$ & $9/13$ & $10/14$ & $11/15$ & $12/16$ & $13/17$ & $14/18$ & $15/19$ & $16/20$ & $21/25$
\\ \cline{2-20}
&$m_h$&--&$10$&$11$&$14$&$17$&$23$&$27$&$31$&$36$&$48$&$53$&$64$&$78$&$87$&$97$&$110$&$122$&$230$
  \\ \hline \hline
\multirow{2}{*}{$w=c=5$}&Code rate &--&--& $1/6$ & $2/7$ & $3/8$ & $4/9$ & $5/10$ & $6/11$ & $7/12$ & $8/13$ & $9/14$ & $10/15$ & $11/16$ & $12/17$ & $13/18$ & $14/19$ & $15/20$ & $20/25$
\\ \cline{2-20}
&$m_h$&--&--&$14$&$22$&$31$&$39$&$49$&$58$&$77$&$91$&$101$&$129$&$150$&$174$&$207$&$232$&$279$&$458$
  \\ \hline \hline
\multirow{2}{*}{$w=c=6$}&Code rate &--&--&--& $1/7$ & $2/8$ & $3/9$ & $4/10$ & $5/11$ & $6/12$ & $7/13$ & $8/14$ & $9/15$ & $10/16$ & $11/17$ & $12/18$ & $13/19$ & $14/20$ &--
\\ \cline{2-20}
&$m_h$&--&--&--&$31$&$42$&$65$&$81$&$97$&$113$&$155$&$185$&$226$&$259$&$301$&$348$&$377$&$446$&--
  \\ \hline
\end{tabular}
\end{table*}

\begin{table}[!t]
\renewcommand{\arraystretch}{1.2}
\caption{Minimum values of $m_h$ for codes with $w=c=3$ and $g=10$ obtained through  the Min-Max algorithm, for several code rates.}
\label{table:Tabg10MinMax}
\centering
\begin{tabular}{|@{\hspace{0.3mm}}c@{\hspace{0.3mm}}|@{\hspace{0.3mm}}c@{\hspace{0.3mm}}|@{\hspace{0.3mm}}c@{\hspace{0.3mm}}|@{\hspace{0.3mm}}c@{\hspace{0.3mm}}|@{\hspace{0.3mm}}c@{\hspace{0.3mm}}|@{\hspace{0.3mm}}c@{\hspace{0.3mm}}|@{\hspace{0.3mm}}c@{\hspace{0.3mm}}|@{\hspace{0.3mm}}c@{\hspace{0.3mm}}|@{\hspace{0.3mm}}c@{\hspace{0.3mm}}|@{\hspace{0.3mm}}c@{\hspace{0.3mm}}|} 
\hline
Code rate& $1/4$ & $2/5$ & $3/6$ & $4/7$ & $5/8$ & $6/9$ & $7/10$ & $8/11$ & $9/12$  
\\ \hline
$m_h$&$11$&$19$&$31$&$53$&$76$&$127$&$222$ &$307$ &$388$
  \\ \hline 
\end{tabular}
\end{table}

\subsection{Numerical simulations of coded transmissions \label{sec:Sims}}

Concerning the \ac{BER} performance of the considered codes, as reasonable and expected, there is a trade-off with their constraint lengths.

We have assessed the performance of our codes through Monte Carlo simulations of \ac{BPSK} modulated transmission over the \ac{AWGN} channel, and compared it with that of some \ac{SC-LDPC-CCs} constructed using the technique described in \cite{Tanner2004}, denoted as Tanner codes. A full size \ac{BP} decoder and a \ac{BP}-based sliding window decoder, both performing $100$ iterations, have been used in the simulations. For each window position, the sliding window decoder takes a decision on a number of target bits equal to $a$.
The $a$ bits that are decided are those in the leftmost $a$ positions inside the decoding window. 
The memory of the sliding window decoder is periodically reset in order to interrupt the catastrophic propagation of some harmful decoding error patterns. The decoder memory reset period has been heuristically optimized for each code.
Let us consider a code with $a=17$, $w=c=3$, $g=8$ and $v_s=646$ (see Table \ref{table:Tabg8MinMax}), denoted as $C_1$. Its exponent matrix is shown in (\ref{eq:Pc1}).
\begin{figure*}
\begin{equation}
\PP_{C_1}=\left[ \small\begin{array}{lllllllllllllllll}
5& 17& 6& 12& 30& 0& 7& 37& 11& 20& 2& 33& 0& 16& 0& 4& 21\\
29& 0& 21& 24& 15& 34& 0& 0& 0& 0& 8& 9& 14& 0& 36& 37& 7\\
0& 28& 0& 0& 0& 32& 30& 29& 0& 21& 0& 0& 37& 36& 2& 0& 0\\
\end{array} \right]
\label{eq:Pc1}
\end{equation}
\end{figure*}

\begin{figure}[!t]
\begin{centering}
\includegraphics[width=85mm,keepaspectratio]{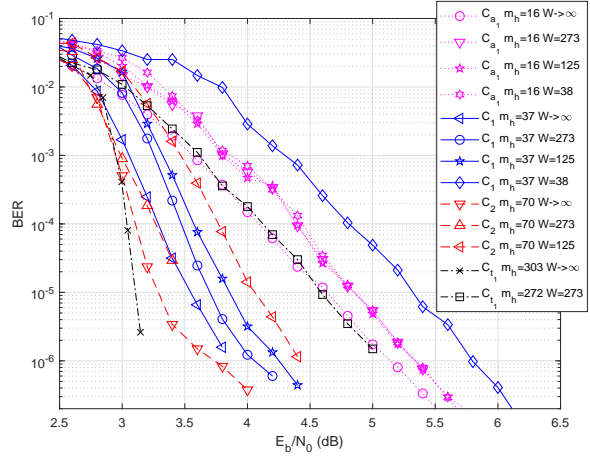}
\caption{Simulated performance of the codes in Table \ref{table:Tabparam}.}
\label{fig:performance}
\par\end{centering}
\end{figure}

We also consider a Tanner code with the same values of $a$, $c$ and $g$, but $v_s=5185$; in addition, by applying the procedures described in Section \ref{sec:Symbolic}, an equivalent code of the Tanner code with $v_s=4641$ has been found and will be called $C_{t_1}$ in the following.
This way, a smaller sliding window can be used also for decoding of the Tanner code. In order to decode $C_1$ with a sliding window decoder \cite{Iyengar2012}, the minimum required window size in blocks is $W=m_h+1=38$, whereas $C_{t1}$ requires at least $W=m_h+1=273$. 
Their performance comparison, obtained by fixing $W=273$, is shown in Fig. \ref{fig:performance}, where the BER is plotted as a function of the signal-to-noise ratio per bit $E_b/N_0$. 
In the figure we also report the performance of the codes simulated using a very large
window.
We notice that $C_{t_1}$ outperforms $C_1$ when $W \rightarrow \infty$, thanks to its larger syndrome former constraint length which likely entails better structural properties, but for small window sizes $C_1$ significantly outperforms $C_{t_1}$. This happens because, as shown in \cite{Lentmaier2011}, values of $W$ roughly $5$ to $10$ times as large as $(m_h + 1)$ result in minimal performance degradation, compared to using the full window size. 
In this case, $W_{C_1}>8m_h$, and $C_1$ approaches its full length performance, whereas $W_{C_{t_1}}=m_h+1$, and $C_{t_1}$ still incurs some loss due to the small window size.

In order to understand the role of the cycles length, another code, noted as $C_2$, has been designed with a heuristic search.
Its exponent matrix is shown in (\ref{eq:Pc2}).
\begin{figure*}[t!]
\begin{equation}
\PP_{C_2}=\left[ \small\begin{array}{lllllllllllllllll}
9	& 59	&30&	44	&0&	55	&0&	0&	65&	0&	21&	0&	58	&37&	24&	0	& 41\\
0&67&26&60	&53	&0	&18	&32	&0	&59	&0&	57&	0	&0	&0	&38	&13\\
5	&0	&0	&0	&9&	55	&70	&42	&27	&14	&43&	16	&68	&57	&56	&41	&0\\
\end{array} \right]
\label{eq:Pc2}
\end{equation}
\end{figure*}

Considering only its first $7$ columns, there are no cycles with length $8$. Finally, the performance of an array code \cite{Fan2000} 
is shown for $W \rightarrow \infty$ and some finite values of $W$.
The parameters of all the considered codes are summarized in Table \ref{table:Tabparam}.

We notice that $C_2$ has a very steep curve in the waterfall region when $W \rightarrow \infty$ but its performance is affected by an error floor.
However, under sliding window decoding with $W=273$, $C_2$ notably outperforms $C_{t_1}$ and $C_1$. Despite its performance is worse than the other ones when $W \rightarrow \infty$, the low value of its constraint length allows it to achieve good performance for very small window sizes, namely $W=38$.

\begin{table}[!t]
\renewcommand{\arraystretch}{1}
\caption{Values of $a$, $c$, $w$, $m_h$, $v_s$ and $g$ of the considered codes with $R=\frac{14}{17}$.}
\label{table:Tabparam}
\centering
\begin{tabular}{|c|c|c|c|c|c|c|c|}
\hline
Code & $a$ & $c$ & $w$ & $m_h$ & $v_s$ & $g$\\ \hline\hline
$C_1$ & $17$ & $3$ & $3$ & $37$ & $646$ & $8$\\ \hline
$C_{t1}$ & $17$ & $3$ & $3$ & $272$ & $4896$ & $8$\\ \hline
$C_{2}$ & $17$ & $3$ & $3$ & $70$ & $1207$ & $6$\\ \hline
$C_{a_1}$ & $17$ & $3$ & $3$ & $16$ & $289$ & $6$\\ \hline
\end{tabular}
\end{table}

\begin{figure}[!t]
\begin{centering}
\includegraphics[width=85mm,keepaspectratio]{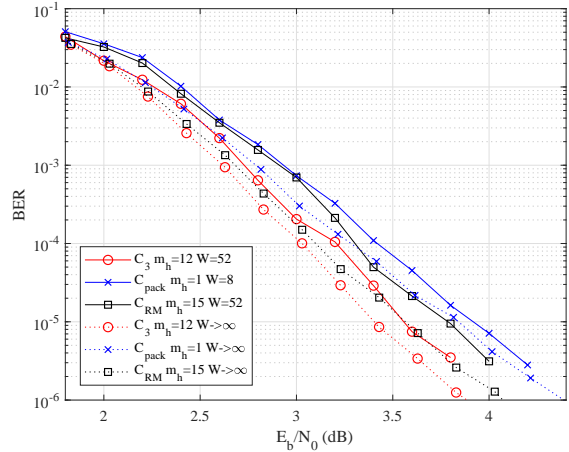}
\caption{BER performance of the codes in Table \ref{table:Tabparam23}.}
\label{fig:perf2}
\par\end{centering}
\end{figure}

In order to provide a comparison with other design approaches, we have used our technique to design a rate-$2/3$ code, denoted as $C_3$, having the following exponent matrix:
\[\PP_{C_3}=\left[\small\begin{array}{cccccccccccc}
12	&10	&9	&7&	3	&0	&0	&5&	0	&4&	1&	10\\
6	&0	&0	&11	&0&	1&	11&	12&	8&	10&	0&	6\\
0	&4	&10	&0&	5&	0	&4&	12&	9&	0&	11&	9\\
12	&9	&5	&0	&1	&9	&10&	0&	3&	8	&12&0
\end{array}\right].
\]
The performance of this code has been compared with that of two codes having similar syndrome former constraint length, but designed through other techniques: an \ac{SC-LDPC-CC} based on packings (see \cite{Zhang2016} for details)
denoted as $C_{\mathrm{pack}}$, and an replicate and mask (R\&M) \ac{SC-LDPC-CC} based on algebraic methods (see \cite{Liu2016})
denoted as $C_{\mathrm{RM}}$. The parameters of the three codes are summarized in Table \ref{table:Tabparam23}.
\begin{table}[!t]
\renewcommand{\arraystretch}{1}
\caption{Values of $a$, $c$, $w$, $m_h$, $v_s$ and $g$ of the considered codes with $R=\frac{2}{3}$.}
\label{table:Tabparam23}
\centering
\begin{tabular}{|c|c|c|c|c|c|c|c|}%
\hline
Code & $a$ & $c$ & $w$ & $m_h$ & $v_s$ & $g$\\ \hline\hline
$C_3$ & $12$ & $4$ & $4$ & $12$ & $156$ & $6$\\ \hline
$C_{\mathrm{pack}}$ & $78$ & $26$ & $4$ & $1$ & $156$ & $6$\\ \hline
$C_{\mathrm{RM}}$ & $12$ & $4$ & $4$ & $15$ & $192$ & $6$\\ \hline
\end{tabular}
\end{table}
Their \ac{BER} performance under sliding window decoding is shown in Fig. \ref{fig:perf2}, where the performance achievable with very large windows ($W \rightarrow \infty$) is also considered as a benchmark.
We notice that the code $C_3$ outperforms the other two codes  with both $W \rightarrow \infty$ and $W_{\mathrm{bits}}=Wa=624$, confirming the effectiveness of our design approach. Note that $W_{C_3}=4(m_h+1)$ and $W_{C_\mathrm{pack}}=4(m_h+1)$, whereas, in order to make the window size in bits comparable, it must be $3<\frac{W_{C_{\mathrm{RM}}}}{m_h+1}<4$. 
$C_{\mathrm{RM}}$ has a value of $m_h$ similar to $C_3$, and its performance is also not far from that of $C_3$.
Instead, $C_{\mathrm{pack}}$ has a very small $m_h$ and apparently this reflects into a significant performance loss with respect to $C_3$.

\section{Conclusion \label{sec:Conclusion}}

We have studied the design of time-invariant \ac{SC-LDPC-CCs} with small constraint length and
free of cycles up to a given length.
By directly designing their parity-check matrix,
 we have been able to obtain codes with smaller constraint length with respect to those designed by unwrapping \ac{QC-LDPC} block codes and, for low values of the girth, the smallest possible ones.
We have also provided lower bounds on the minimum constraint length which is needed to
achieve codes with a fixed girth, and shown through new design and search methods that practical codes exist which are able to approach or even reach these bounds.

\begin{appendices}

\section{}

Let us describe the IP optimization model we use. 
As inputs, it takes a big enough penalty $M$, all the entries of an exponent matrix $\PP$, a positive integer $p \rightarrow \infty$ to represent the maximum allowed exponent in $\PP$, and the set of relatively prime numbers to $p$ (this set has cardinality $\phi(p)$, where $\phi$ is the Euler function) and it finds Min-Max of the elements of all the matrices $\PP_t$'s, where $\PP_t$ is the transformed exponent matrix obtained by applying Lemmas \ref{lem:sumequi} and \ref{lem:proequi} on $\PP$. Note that for specific parameters of $w$, $c$, $a$ and $g$, we picked the exponent matrix $\PP$ from \cite{Tasdighi2016,BocharovaHugJohannessonEtAl2012,Tasdighi2016a}, where $\PP$ has the smallest possible circulant size $p$. Thus, if $\PP^*_t$ is one of the optimal transformed exponent matrices, our model explicitly finds the linear transformations based on Lemmas \ref{lem:sumequi} and \ref{lem:proequi} and, by applying them on $\PP$, we achieve new instances of $\PP^*_t$. Furthermore, the model finds the maximum syndrome former memory order $m_h$ in $\PP^*_t$ as output; the final $m_h$ is the minimum possible $m_h$ of all the transformed matrices $\PP_t$.

In the following list, we enumerate the steps of our model:

\begin{align*}\begin{array}{ll}
\textbf{1.}&\text{minimize}\quad Z=\sum_{i=0}^{a-1}\sum_{j=0}^{c-1} x_{ij}  \\
&\;\text{s.t.}\;\;\\
\textbf{2.}&\;\;\;b_{ij}=\left(\sum_{g=1}^{\phi(p)} k_g T_g\right)p_{ij}+r_i+c_j \;\;\;
i\in\lbrace 0,\cdots , a-1 \rbrace \;\&\; j\in\lbrace 0,\cdots , c-1 \rbrace \\
\textbf{3.}&\;\;\;\sum_{g=1}^{\phi(p)} k_{g} = 1\\
\textbf{4.}&\;\;\;p\psi_{ij}\leq b_{ij}\\
\textbf{5.}&\;\;\;b_{ij}+0.5\leq \left(1+\psi_{ij}\right)p\\
\textbf{6.}&\;\;\;d_{ij}= b_{ij}-p\psi_{ij}\\
\textbf{7.}&\;\;\;d_{mn}\leq d_{ij}+M\left(1-y_{ij}\right)\;\;\;\;\left(m,n\right)\neq\left(i,j\right)\\
\textbf{8.}&\;\;\;\sum_{i=0}^{a-1}\sum_{j=0}^{c-1} y_{ij}=1\\
\textbf{9.}&\;\;\; x_{ij}\leq My_{ij} \\
\textbf{10.}&\;\;\; d_{ij}\leq x_{ij}+M\left(1-y_{ij}\right) \\
\textbf{11.}&\;\;\;k_g,y_{ij}\in \lbrace 0,1\rbrace,\;0\leq r_i,c_j<p,\;\text{and}\;r_{i},c_{j},\psi_{ij},d_{ij},x_{ij}\;\text{are integers.}  
\end{array}
\end{align*}
A brief description of the steps of the model is as follows.
Each element $p_{ij}$ of $\PP$ is transformed to $b_{ij}$ by multiplying it to a relatively prime number $T_g$, as well as by adding two decision variables $r_i$ and $c_j$ (Constraint 2). Constraint 3 indicates that just one of the relatively prime number to $p$ could be selected. Constraints 4 and 5 determine the quotient $\psi_{ij}$ of element $b_{ij}$ divided by $p$. Constraint 6 is the residual of subtracting $p\psi_{ij}$ from $b_{ij}$. The two constraints 7 and 8 are used to detect the maximum element of the transformed exponent matrix modulo $p$, where, $y_{ij}$'s are identification binary variables. Clearly, just one of these variables can be chosen. Variables $x_{ij}$'s in constraints 9 and 10 are created in such a way that just one of them is greater than zero, which is the maximum among all of the elements in $\PP^*_t$. This output element is considered to be Min-Max or our desired $m_h$.

\section{}

Let us compute complexity of performing exhaustive analyses of codes with fixed parameters by using their binary or symbolic matrix representation.

\subsection{Binary matrix}

Let us consider $\HH_s^T$ and, for the sake of simplicity, let us suppose that $L_h$ is an integer multiple  of $c$; so, $\HH_s^T$ has size $L_h \times a$. Considering that any column has weight $w$, and that all columns must be different in order to satisfy the \ac{r.c.c.}, we can enumerate the number of possible matrices as
\begin{equation}
N=\binom{\binom{L_h}{w}}{a}.
\end{equation}
The space of possible matrices can be further reduced if we consider that any code has an equivalent code with at least a $1$ in the first position of the first column of $\HH_s^T$, yielding
\begin{equation}
N={\binom{L_h-1}{w-1}}{\binom{\binom{L_h}{w}}{a-1}}.
\end{equation}
Furthermore, any code has an equivalent code with at least a $1$ in any of the first $c$ positions of $\HH_s^T$. Therefore, we have
\begin{equation}
N={\binom{L_h-1}{w-1}}\binom{\sum_{i=1}^c{\binom{L_h-i}{w-i}}}{a-1}.
\end{equation}

Finally, any code has $a!$ equivalent codes obtained by permuting the columns of $\HH_s^T$. This means that the number of possible matrices can be eventually reduced to
\begin{equation}
N=\binom{\sum_{j=1}^{\binom{L_h-1}{w-1}}{\left[\sum_{i=1}^c{\binom{L_h-i}{w-i}}-j\right]}} { a-1}.
\end{equation}

\subsection{Symbolic matrix}

Let us consider an exponent matrix $\PP$ with size $c \times a$ such that its $(i, j)$-th element $p_{i,j} \in [0,m_h]$. In general, $N=(m_h+1)^{ac}$ matrices should be tested. However, if we consider that all the columns of a symbolic matrix must be different, we obtain $N=\binom{(m_h+1)^c}{a}$ possible matrices. By using Lemmas \ref{lem:permequi} and \ref{lem:sumequi} it is possible to show that any code has an equivalent code with at least a null value in any column of $\PP$. Since there can be $m_h^c$ possible columns of $\PP$ without a null symbol, out of the total $(m_h+1)^c$ possible columns, we obtain
\begin{equation}
N=\binom{(m_h+1)^c-m_h^c}{a}.
\end{equation}

Moreover, due to Lemma \ref{lem:permequi}, any code has an equivalent code with at least one null entry in $p_{1,1}$, reducing again the space of possible matrices to
\begin{equation}
N=(m_h+1)^{c-1}\binom{(m_h+1)^c-m_h^c}{a-1}.
\end{equation}

Finally, exploiting again Lemma \ref{lem:permequi}, we can consider any $n$-tuple of columns only once (without taking into account its permutations), thus obtaining
\begin{equation}
N=\binom{\sum_{j=1}^{(m_h+1)^{c-1}}[(m_h+1)^c-m_h^c-j] }{a-1}.
\end{equation}

\end{appendices}

\bibliographystyle{IEEEtran}
\bibliography{Archive}
\end{document}